%% file: neurips_2026.tex
\documentclass{article}

\usepackage{natbib}
\usepackage{needspace}   

\usepackage[main, final]{neurips_2026}

\usepackage[utf8]{inputenc} 
\usepackage[T1]{fontenc}    
\usepackage{hyperref}       
\usepackage{url}            
\usepackage{booktabs}       
\usepackage{amsfonts}       
\usepackage{nicefrac}       
\usepackage{microtype}      
\usepackage{xcolor}         
\usepackage{wrapfig}
\usepackage{placeins}      
\usepackage{microtype}
\usepackage{graphicx}
\usepackage{subcaption}
\usepackage{booktabs} 
\usepackage{multirow}   
\usepackage[table]{xcolor} 

\usepackage{hyperref}
\usepackage{multirow}

\usepackage{amssymb}
\usepackage{mathtools}
\usepackage{amsthm}

\providecommand{\grp}[2]{\rowcolor[gray]{.94}\multicolumn{#1}{@{}l}{\textbf{#2}}}
\providecommand{\hd}[1]{\textbf{#1}}
\providecommand{\up}{$\uparrow$}
\providecommand{\dn}{$\downarrow$}
\providecommand{\pmsd}[1]{{\scriptsize\color{gray}$\pm$#1}}
\providecommand{\ERt}{ER$_{20^{\circ}}$}
\providecommand{\Ft}{F$_{20^{\circ}}$}
\providecommand{\LEcd}{LE$_{\text{CD}}$}
\providecommand{\LRcd}{LR$_{\text{CD}}$}

\title{\textsc{AudioSphere}: Towards Self-Supervised Spatial Audio Representation Models}

\author{%
  Goksenin Yuksel \\
  Donders Institute, Radboud University \\
  Nijmegen, the Netherlands \\
  \texttt{goksenin.yuksel@donders.ru.nl}\\
  \And
  Marcel van Gerven \\
  Donders Institute, Radboud University \\
  Nijmegen, the Netherlands \\
  \texttt{marcel.vangerven@donders.ru.nl} \\
  \AND
  Kiki van der Heijden\\
  Donders Institute, Radboud University, Nijmegen, the Netherlands \\
  Mortimer B Zuckerman Institute, Columbia University, New York, United States \\ 
  \texttt{kiki.vanderheijden@donders.ru.nl} \\
}

\begin{document}

\maketitle

\begin{abstract}

Recent audio embedding models learn representations that transfer across clip- and frame-level tasks, yet they operate almost exclusively on single-channel signals and are blind to spatial structure. Spatial audio models capture that structure but rely on clip-level supervision or contrastive objectives, yielding embeddings unsuitable for frame-level tasks like sound event detection and localization (SELD). We argue this dichotomy is not fundamental: patch-level spatial self-supervision can jointly learn acoustic and spatial features at both temporal granularities. We introduce AudioSphere, a multi-channel masked autoencoder trained on Ambisonics recordings that reconstructs both spectral content and directional cues via a joint spatiotemporal masking strategy that masks both time-frequency patches and intensity vectors. AudioSphere supports frame-level predictions while producing strong clip-level representations via aggregation. Alongside the model, we release RealSELD, a standardized benchmark for evaluating frozen backbones on real-world multi-channel SELD. AudioSphere enables, for the first time, SELD with a frozen self-supervised backbone, while matching or exceeding prior self-supervised audio embedding models on clip-level classification tasks spanning environmental sound, speech, and music, and on frame-level sound event detection.\footnote{All code and data are available at \url{https://github.com/labhamlet/AudioSphere} and \url{https://huggingface.co/labhamlet}.}

\end{abstract}

\section{Introduction}

Despite the complexity and diversity of everyday sound scenes, human listeners effortlessly perceive both \emph{what} they hear and \emph{where} it comes from. Audio embedding models aspire to this breadth~\citep{hear, superb}, and recent self-supervised learning (SSL) approaches achieve strong performance on conventional benchmarks with no fine-tuning~\citep{beats, wav2vec2, mwmae}. Yet, state-of-the-art models are trained almost exclusively on single-channel corpora such as AudioSet~\citep{audioset} and LibriSpeech~\citep{librispeech}. This renders them essentially ``spatially blind.'' This omission is consequential: in humans, spatial cues are central to auditory scene analysis, aiding the separation of overlapping sound waves into distinct auditory objects~\citep{bregman1984auditory, heijden2019cortical}. Consequently, the absence of spatial information in learned embeddings not only precludes sound localization but also limits their ability to parse complex acoustic scenes.

While SSL has produced embeddings that transfer across a wide range of monaural audio tasks, a fundamental divergence persists in current architectures. Monaural models achieve remarkable performance by scaling to vast unlabeled corpora, but ignore the spatial dimension~\citep{beats, wav2vec2, mwmae}. Conversely, spatial models like BAT~\citep{spatialast} and ELSA~\citep{elsa} capture directional cues but remain limited to clip level supervised labels or contrastive spatial metadata. Labeled spatial supervision has three consequences. It restricts training to annotated real-world scenes or to simulated scenes with known geometry. Unless such models are trained at very large scale, the resulting representations are optimized for a particular task or dataset and transfer poorly. And their clip-level objectives are not designed to learn the frame-resolved representations required for tracking moving sources. 

We address this limitation by pre-training an audio representation model on multi-channel audio recordings using spatial-acoustic patch-level self-supervision. In this work, we adopt the Ambisonics spatial audio format~\citep{ambisonics} for learning spatial audio embeddings. Ambisonics provides a compact, geometry-independent encoding of the spatial sound field, decoupling the representation from the recording hardware~\citep{ambisonics}. Moreover, any multi-channel recording can be projected into this canonical format via Ambisonics Signal Matching (ASM)~\citep{acm_ref}, a signal-processing technique that maps signals from an arbitrary microphone array to Ambisonics. This projection preserves the spatial cues of the original sound field while yielding a fixed-dimensional input, independent of the number or arrangement of microphones.

To this end, we introduce \textsc{AudioSphere}, a multi-channel masked autoencoder that learns joint acoustic and spatial representations. \textsc{AudioSphere} reconstructs both log-mel spectrograms and active intensity vectors through a tube-masking strategy that jointly masks spectral and spatial patches, forcing the encoder to internalize spatial information alongside the acoustic features required for clip-level classification and frame-level event detection. To train \textsc{AudioSphere} at scale, we develop a simulation pipeline that generates 70,000 high-fidelity Ambisonic room impulse responses (ARIRs) from Matterport3D~\citep{Matterport3D} environments, paired with an in-batch spatial sampling strategy that increases the effective batch size 16-fold while amortizing computational costs.

The main contributions of our work are as follows.
(1)~\textsc{AudioSphere} learns joint spatial and acoustic embeddings through self-supervised training, unlocking real-world SELD with a single model.
With only shallow MLP probing on a frozen backbone, it outperforms fully supervised baselines and prior spatial models on static scenes and transfers competitively to dynamic environments. The same embeddings also serve as strong acoustic representations, transferring to environmental-sound, speech, and music classification, as well as frame-level sound event detection.
(2)~To support reproducible evaluation of spatial audio representations, we release \textsc{RealSELD}, the first evaluation kit for real-world SELD, which adapts five established real-world SELD datasets into a frozen evaluation format.
(3)~By using Ambisonics as a canonical representation, \textsc{AudioSphere} processes tetrahedral, planar, and spherical microphone arrays without retraining; on real-world recordings from a planar array (RSL2019-dev), the ASM~+~\textsc{AudioSphere} pipeline achieves a median direction-of-arrival error of $1.9^\circ$, compared with $5.6^\circ$ for MUSIC and $13.1^\circ$ for the Binaural Signal Matching (BSM) ~+~Spatial-AST (BAT) pipeline.

\section{Related Work}
\label{sec:related-work}

\textbf{Self-supervised audio representation learning.} The dominant paradigm for learning transferable audio representations is masked spectrogram modeling with Transformer backbones. Such methods include SSAST~\citep{gong2022ssast}, MSM-MAE~\citep{masked-modelling}, MaskSpec~\citep{MaskSpec}, MAE-AST~\citep{mae-ast}, and MW-MAE~\citep{mwmae}, which achieves strong performance on the HEAR benchmark through multi-window local-global attention in the decoder. SSAM~\citep{audiomamba} replaces the Transformer encoder with a Mamba backbone~\citep{mamba} and attains comparable results. Recently, Dasheng~\citep{dasheng} demonstrated the impact of massive data scaling, achieving state-of-the-art performance on the HEAR benchmark by increasing the pre-training volume from the standard ${\sim}$5.8k hours to 272k hours. Beyond reconstruction-based objectives, other methods operate on latent representations. BEATs~\citep{beats} masks latent embeddings produced by an acoustic tokenizer, while ATST-Frame and ATST-Clip~\citep{atst} utilize a student-teacher self-distillation paradigm to reconstruct frame-level latent embeddings from masked spectrogram views. Contrastive approaches such as COLA~\citep{COLA} dispense with masking entirely. A parallel line of work targets speech specifically; models such as Wav2Vec2~\citep{wav2vec2}, HuBERT~\citep{hubert}, and WavLM~\citep{wavlm} achieve strong speech performance but transfer less reliably to non-speech tasks~\citep{hear}. Critically, all of the above operate on single-channel audio and learn no spatial representation of the sound scene.

\textbf{Spatial audio representation learning.} While spatial audio has recently gained traction, existing models are typically trained in a supervised manner, relying on synthetic labels from simulated sound scenes. Supervised approaches, such as BAT~\citep{spatialast} and ELSA~\citep{elsa}, rely on labels or natural-language metadata, thereby limiting their training to fully synthetic datasets. While models optimized exclusively for regressing to synthetic spatial labels \citep{spatial-libri} achieve localization, they do not jointly learn the acoustic classes and thus fail to capture the underlying acoustic properties of the audio signal. Furthermore, because these frameworks rely on static, clip-level supervision, their objectives are not designed to learn the frame-resolved representations required for moving-source SELD. To mitigate these supervision and temporal granularity bottlenecks, w2v-SELD~\citep{w2vseld} recently explored learning representations from unlabeled Ambisonics data. w2v-SELD operates entirely in the waveform domain using a wav2vec2-based objective strictly tailored for SELD. Consequently, it does not address transfer to tasks beyond SELD, and its generalization capacity is severely bottlenecked by a pre-training corpus limited to 200 hours. Recent audio-LLM frameworks, such as Sci-Phi~\citep{sciphi} and Whisper-based variants~\citep{tang2024largelanguagemodelsunderstand}, offer natural-language interfaces for static scenes on clip-level tasks, but their audio embeddings have not been shown to transfer to spatial and acoustic downstream tasks. They effectively yield a natural language interface for spatial sound scene understanding rather than a spatial-acoustic representation model.

\textsc{AudioSphere} differs from the state-of-the-art in three key dimensions: (i) It learns joint spatial-acoustic features via a frame-level reconstruction objective, requiring zero ground-truth labels; (ii) Unlike task-specific encoders, \textsc{AudioSphere} produces embeddings that excel at both acoustic and spatial downstream tasks through a shallow MLP probe on frozen features; (iii) As a self-contained front-end, it provides a ``plug-and-play'' spatial-acoustic layer for any downstream architecture, including existing audio-LLM pipelines.

\section{Learning Self-Supervised Spatial Audio Representations}

\subsection{First-order Ambisonics}
\label{sec:foa}

We assume a far-field acoustic scenario in which the sound field is composed of a distribution of plane waves. The spatio-temporal amplitude of the sound field, denoted as $s(t, \Omega)$, depends on time $t$ and the direction of arrival $\Omega = (\theta, \phi)$, where $\theta \in [0, 2\pi)$ denotes azimuth and $\phi \in [-{\pi}/{2}, {\pi}/{2}]$ denotes elevation.
 
Ambisonics relies on the decomposition of the spherical sound field into orthogonal spatial basis functions known as Spherical Harmonics (SH). The sound field can be reconstructed from its Ambisonic coefficients via:
\begin{equation}
s(t, \Omega) = \sum^{\infty}_{l=0} \sum^{l}_{m=-l} B_{lm}(t) Y_{lm}(\Omega)
\end{equation}
where $B_{lm}(t)$ are the Ambisonic coefficients of order $l$ and degree $m$. In practice, the summation is truncated at a maximum order $L$, resulting in $K = (L+1)^2$ channels. The basis functions $Y_{lm}(\theta, \phi)$ are Real Spherical Harmonics defined using associated Legendre functions $P_l^{|m|}(\sin\phi)$. Here, we utilize the  N3D normalization and ACN channel ordering (see Appendix~\ref{appendix:sh} for full definitions). Ambisonics, therefore, serves as a canonical representation of the 3D sound field, providing a reusable format for encoding spatial audio from any microphone geometry.

\textbf{Ambisonics Signal Matching (ASM). }To enable \textsc{AudioSphere} to process diverse microphone arrays, we adopt Ambisonics Signal Matching (ASM)~\citep{acm_ref, ref_ben}, which generalizes traditional Ambisonic encoding to arbitrary configurations via Tikhonov regularization~\citep{tik}. While parallel work explores deep neural networks for this mapping~\citep{heikkinen2026omnidirectionalneuralambisonicsencoding}, we opt for the established ASM framework given its established effectiveness. Our pipeline is modular; neural format conversion techniques can be integrated as they advance. We use the \texttt{SHroom}~\citep{shroom} for all ASM computations. ASM is a fixed linear front end with its filters precomputed from the array geometry and contains no trainable parameters, but the mapping is differentiable with respect to its input.

\begin{figure}[!t]
    \centering
    \includegraphics[width=\linewidth]{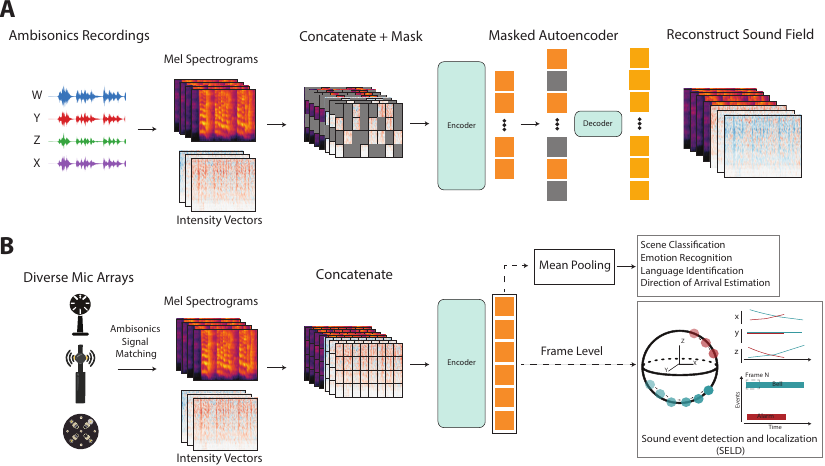}
  \caption{\textbf{(A) Pre-Training:} We employ the Masked Autoencoder (MAE) approach to learn spatial audio representations. Ambisonics recordings are converted into mel spectrograms and Intensity Vectors, concatenated, and masked. The Encoder processes unmasked patches and the Decoder reconstructs the original full sound field.
    \textbf{(B) Inference:} The pre-trained Encoder can be flexibly applied to diverse microphone arrays. Diverse input signals are standardized via Ambisonics Signal Matching and passed through the Encoder to extract frame-level and scene-level Ambisonics embeddings.
  }
  \label{fig:architecture}
\end{figure}

\subsection{Dataset Generation}
We generate training data in three stages: (i) high-quality simulation of room acoustics via Ambisonic room impulse responses (ARIRs), (ii) stochastic scene composition, and (iii) online mixing with AudioSet sound clips.
 
\textbf{Simulating acoustic scenes.}
Large-scale, natively recorded Ambisonics datasets remain scarce: the largest available corpus contains 246 hours of audio~\citep{kim2025visage,yt360}, compared to the standard ${\sim}5.8$K hours of single-channel sound clips in AudioSet~\citep{audioset}. Here, we used the ${\sim}4.8$\,K hours (${\sim}1.74$M clips) currently available from AudioSet. To bridge this gap, we leverage high-resolution 3D meshes from Matterport3D~\citep{Matterport3D} and the Monte Carlo ray-tracing simulator from SoundSpaces~2.0~\citep{chen22soundspaces2} to generate Ambisonic room impulse responses (ARIRs) that encode both room acoustics and first-order spatial cues in realistic scenarios. 

\textbf{Scene composition.}
We randomly selected 70 MatterPort3D houses with a sufficient mesh quality. For each house we simulate 1{,}000 scenes, each comprising a randomly sampled listener position, a sound source (distance $\in [1.5, 5]$\,m; azimuth $\in [0, 2\pi)$; elevation $\in [-{\pi}/{2}, {\pi}/{2}$], and one or more noise sources. Half of the scenes use localized noise at a single position; the other half use diffuse noise from three to five randomly placed sources. For each configuration we render $\mathrm{ARIR}(s, r, \theta)$ from source to listener and $\mathrm{ARIR}(\phi_i, r, \theta)$ from each noise source to listener, yielding 70{,}000 ARIR sets (see Appendix~\ref{appendix:brir-spec} for full parameters). We restrict our pre-training simulations to static sources, as moving sources require sampling hundreds of RIRs per trajectory and thus scale poorly for self-supervised pre-training. We note, however, that because \textsc{AudioSphere} learns spatial cues at the patch level, it can localize moving sources during inference (empirically validated in Section~\ref{ref:moving}).
 
\textbf{Online mixing.}
During training, each clip from the AudioSet training set~\citep{audioset} (${\sim}1.74$M clips, ${\sim}4.8$K hours) is randomly paired with a WHAM! noise clip~\citep{Maciejewski2020WHAMR} and trimmed to 10\,s with a 200\,ms linear fade-in/fade-out. The target signal $T$ is obtained by convolving the AudioSet clip with $\mathrm{ARIR}(s,r,\theta)$, and each noise clip produces $N_i$ via convolution with $\mathrm{ARIR}(\phi_i,r,\theta)$. For diffuse scenes the noise field is $D = \sum_{i=1}^{M} N_i$. The final scene is:
\begin{equation}
    S = \begin{cases} T + bN & \text{(localized noise)} \\ T + bD & \text{(diffuse noise)} \end{cases}
\end{equation}
where $b$ sets the signal-to-noise ratio, sampled uniformly from $[+5, +40]$\,dB. Although AudioSet clips may already contain reverberation and background noise, we prioritize acoustic diversity following established practice~\citep{spatialast, elsa, politis_2022_6406873}.

\subsection{Model Training}
\label{sec:model_training}

\textbf{Input features.}
We transformed the channels of each sound scene (i.e., the waveforms) into log-scale mel spectrograms using 128 mel filters in the frequency range of 50-16000 Hz with a 25 ms Hanning window and 10 ms hop length, resulting in spectrograms of dimension $1001 \times 128$. We added zero padding to achieve a dimension of $1024 \times 128$.  For \textsc{AudioSphere}, following \citet{elsa, ambisonic_IVs}, we  extracted intensity vectors utilizing the equation below:
 \begin{equation}
     \label{eq:iv}
     \boldsymbol{I}_\text{active}(t,f) = \Re \left[ A_\text{0,0}^{*}(t,f) \begin{pmatrix}A_\text{1,-1}(t,f)\\ A_\text{1,0}(t,f)\\ A_\text{1,1}(t,f) \end{pmatrix} \right], 
 \end{equation}
 
 where $A_{n,m}$ are the $n^\textrm{th}$ and $m^\textrm{th}$ order and mode of the ambisonics signal corresponding to its omnidirectional ($W$) and three dipole $(Y,Z,X)$ components,  and $(\cdot)^*$ denotes complex conjugation. IVs are scaled to unit norm. These two targets address acoustic and spatial aspects of the Ambisonics sound field. The log-mel spectrograms of the $W$, $X$, $Y$, $Z$ channels encode the acoustic content of the scene, while the active intensity vectors, the real part of the cross-spectrum between $W$ and the directional components, integrate the relative phase and magnitude relations between channels into a physically interpretable directional vector. We concatenated mel spectrograms and intensity vectors, resulting in input $x = [x_\text{mel}, IVs]$.

\textbf{Patching and masking.}
A single convolutional layer partitions each multi-channel spectrogram into $N$ non-overlapping patches $\{P_i\}_{i=1}^{N}$, $P_i \in \mathbb{R}^{7 \times 8 \times 16}$, and projects each into a $d$-dimensional embedding $E_i \in \mathbb{R}^{d}$ with $d{=}768$. Fixed 2D sinusoidal positional embeddings are added. We employ \emph{tube masking}: 80\% of time-frequency patch locations are selected uniformly at random, and the corresponding patches are masked across all input channels simultaneously. Masked embeddings are replaced with a shared learnable mask token during decoding.

Tube masking couples spatial and acoustic learning. Because the Ambisonics and intensity-vector channels share one time--frequency grid, masking a location in only some channels would let the model copy it from the unmasked channels at the same location, which impedes learning joint acoustic-spatial representations. Masking all channels at once forces the encoder to infer the directional channels and intensity vectors from the surrounding spatial context. Appendix~\ref{appendix:cleanratio} (Table~\ref{tab:ablations}) compares tube masking with per-channel masking and cross-channel reconstruction targets, and varies masking pattern and ratio. The results suggest that tube masking outperforms 
per-channel masking on both an acoustic probe and a spatial one, whereas a random masking pattern with a high ratio is beneficial, so we use random tube masking at 80\%.
 
\textbf{Encoder.}
The unmasked embeddings are processed by a 12-layer ViT-Base encoder~\citep{dosovitskiy2020vit} (hidden dimension = 768, 12 attention heads, MLP dimension = 3072), producing representations $\{O_i\}_{i=1}^{N'}$ where $N'$ is the number of visible patches.
 
\textbf{Decoder.}
The decoder takes as input the unmasked encoder outputs $\{O_i\}$ and the learnable mask tokens $\{M_i\}$, with $M_i \in \mathbb{R}^{768}$, augmented with fixed sinusoidal positional embeddings. The decoder consists of eight Transformer layers with local-global attention~\citep{mwmae}, using window sizes $[2, 5, 10, 25, 50, 100, 0, 0]$ where 0 denotes global (full) attention. A linear projection head reconstructs the four-channel spectrogram and three IVs from the decoder output ($768 \to 7 \times 8 \times 16$).
 
\textbf{In-batch sampling.}
As the online mixing of naturalistic
acoustic scenes is computationally expensive, we introduced a random in-batch sampling
procedure to increase the effective batch size in a computationally efficient manner. We randomly sampled 16 partially overlapping 2-second segments, yielding 16 samples of dimension 200 × 128. This increases the nominal batch size of 96 to an effective batch size of 1536.
 
\section{Downstream Evaluation}

\begin{table}[!t]
\renewcommand{\arraystretch}{1.2}
\setlength{\tabcolsep}{3pt}
\centering
\scriptsize
\caption{\textbf{The \textsc{RealSELD} datasets}. \emph{Impulse response}: real room impulse responses convolved with isolated sound events (Real) or direct recordings (--). }
\label{tab:RealSELD-datasets}
\begin{tabular}{@{}lllll@{}}
\toprule
\textbf{Name} & \textbf{Motion} & \textbf{Impulse Response} & \textbf{Max Overlap} & \textbf{Noise} \\ \midrule
TUT-2018 Real (ov1, 2, 3)~\citep{seldnet} & Static & Real & 1, 2, 3 & Ambient 30\,dB  \\
TAU-2019~\citep{tau2019} & Static & Real & 2 & Ambient 30\,dB\\
TAU-2020~\citep{tau2020} & Dynamic & Real & 2 & Ambient [30-6\,dB] \\
TAU-2021~\citep{tau2021} & Dynamic & Real & 3 & Dynamic + Ambient\\
STARSS23~\citep{STARSS23} & Dynamic & -- & 6 & Dynamic + Ambient \\ \bottomrule
\end{tabular}
\end{table}

We aim to answer four questions using the pre-trained, frozen \textsc{AudioSphere} encoder. (i)~\emph{Does spatial pre-training transfer to monaural audio understanding?} We test this with HEAR~\citep{hear}, a suite of classification and detection tasks consisting of single-channel audio. (ii)~\emph{Do the learned embeddings carry frame-resolved spatial information that supports real-world SELD, in static and in dynamic scenes?} We test this with \textsc{RealSELD}, our evaluation kit of real-world multi-channel data (Section~\ref{ref:moving}). (iii)~\emph{Does all of this hold across microphone geometries?} We test this on tetrahedral and planar arrays. (iv)~\emph{Does the encoder rely on spatial cues to understand spatial scenes?} We test this by removing or corrupting  spatial information at pre-training or inference time.

\textbf{Evaluating the transfer of spatial-acoustic audio representations to monaural tasks.} 
We use the subset of HEAR from \citet{mwmae}: \emph{acoustic events and scenes}, i.e., tagging (FSD50K), classifying environmental sounds (ESC-50), and counting speakers (LibriCount); \emph{speech}, i.e., emotion recognition (CREMA-D), spoken-language identification (VoxLingua107), and keyword spotting (Speech Commands); and \emph{music}, i.e., pitch classification (NSynth) and percussion instrument, stroke, and tonic classification (Beijing Opera, Mridangam). Because this subset contains no timestamp-level task, we add the HEAR sound event detection task DCASE-2016 Task~2~\citep{Mesaros2018_TASLP}. Appendix~\ref{appendix:tasks} lists all tasks. As a monaural reference we include \textsc{AudioSphere}-Clean, trained on AudioSet with the MW-MAE framework plus our in-batch sampling, and we ablate the ratio of clean (single-channel) to simulated (four-channel) pre-training data. Single-channel audio is duplicated across channels when mixed with four-channel data, following \citet{elsa}.

\textbf{Evaluating the transfer to SELD tasks in real-world scenes.}
To test in a standardized manner how well audio representation models perform SELD in real-world scenes, we introduce \textsc{RealSELD}. \textsc{RealSELD} embeds five established real-world SELD datasets (seven evaluation tasks) in the HEAR evaluation format. The datasets differ in difficulty and realism (Table~\ref{tab:RealSELD-datasets}). TUT-2018 Real~\citep{seldnet} and TAU-2019~\citep{tau2019} contain \emph{static} scenes with one to three simultaneous sources and ambient noise at 30\,dB SNR. TAU-2020 and TAU-2021 introduce \emph{moving} sources and more complex background noise. STARSS23 contains real recordings of natural indoor scenes with up to six simultaneous moving sources and complex background noise. The TUT and TAU datasets are constructed by convolving isolated sound events with real room impulse responses recorded in diverse and challenging environments, including underground bomb shelters, gym spaces, classrooms, glass-walled meeting rooms, and lecture halls, while STARSS23 is recorded directly.

The SELD task is formulated using the ACCDOA framework~\citep{shimada2021accdoa} in which the probe predicts, per frame and per class, a vector whose direction is the direction-of-arrival (DoA) and whose length indicates whether the class is active. We report location-dependent metrics for dynamic sound scenes: Error rate ER$_{20^{\circ}}$ and F-score F$_{20^{\circ}}$, which count a detection as correct only if the class is right and the predicted direction is within $20^{\circ}$ of the reference; class-aware localization: localization error LE$_{CD}$ (angular distance of correct-class predictions) and localization recall LR$_{CD}$ (how often a direction is produced for a reference event); and location-independent metrics for static sources following the DCASE convention (Appendix~\ref{appendix:accdoa}). Note that to resolve the mismatch between the 80\,ms temporal resolution of the \textsc{AudioSphere} embeddings and the 100\,ms label resolution of the dynamic RealSELD datasets, we interpolated the embeddings to the 100\,ms label resolution using average pooling.

\textbf{Evaluating benefit of pre-training for dataset-specific fine-tuning.}
\label{sec:starss23}
Although a pre-trained audio representation model may perform suboptimally on a task data distribution that diverges strongly from the training distribution in comparison with a supervised model fine-tuned extensively on this particular distribution, the characteristics of the learned representations may nevertheless aid the model's transfer to such a specific task data distribution by minimizing the need for extensive fine-tuning. To assess the transferability of \textsc{AudioSphere}'s embeddings, we compared three training scenarios. \emph{Scratch}: training the \textsc{AudioSphere} architecture on the target dataset only, instead of pre-training on our training dataset. \emph{Frozen probing}: evaluating the frozen pre-trained model with a trainable downstream head on the target dataset, fitting only the head, which is either the shallow MLP probe used throughout or a lightweight temporal head from the SELDnet baseline. \emph{Fine-tuning}: a two-step approach in which the model is first pre-trained on our training dataset, followed by joint optimization of the model and the downstream head on the target dataset.  On STARSS23~\citep{STARSS23} (we use the extended version of STARSS23 dataset as the DCASE 2023 Baseline) we compared training \textsc{AudioSphere} from scratch versus fine-tuning the pretrained \textsc{AudioSphere}: to train from scratch, we tested four different learning rates $\{10^{-3}, 5{\times}10^{-4}, 2{\times}10^{-4}, 10^{-4}\}$, and for fine-tuning we used a learning rate of $10^{-4}$.

\textbf{Evaluating the generalization across microphone array geometries.} We evaluate real-world recordings from various arrays: the tetrahedral arrays of TAU-2019, TAU-2020, TAU-2021 and STARSS23 and the planar ReSpeaker~2.0 array of RSL2019-dev~\citep{rsl2019}, all encoded to first-order Ambisonics with ASM. RSL2019-dev is converted to the HEAR format with stratified train/validation/test splits (50/25/25\%) by source location; we re-implement MUSIC~\citep{MUSIC} on the held-out split and evaluate the BAT (Spatial-AST) encoder on the same split after mapping the recordings to its binaural input with Binaural Signal Matching (BSM)~\citep{gayer2025ambisonicsencoderwearablearray} (Appendix~\ref{appendix:bsm}).

\section{Results}

\subsection{\textsc{AudioSphere} Transfers to Clip-Level Classification and Frame-Level Detection on HEAR}
\label{sec:res_hear}

\begin{wraptable}{r}{0.44\textwidth}
\vspace{-12pt}
\centering
\scriptsize
\setlength{\tabcolsep}{4pt}
\renewcommand{\arraystretch}{1.1}
\caption{Aggregate HEAR score (mean over the 11 tasks of Appendix Table~\ref{tab:tasks_overall}; per-task results in Appendix Table~\ref{tab:dry_audio_benchmarks}), $\pm$\,SD over three downstream seeds. Grey: supervised. \textbf{AS}: AudioSet.}
\label{tab:hear_summary}
\begin{tabular}{@{}llc@{}}
\toprule
\textbf{Model} & \textbf{Corpus} & \textbf{HEAR} \\
\midrule
BEATs       & AS  & 76.6 \\
SSAM        & AS  & 79.6 \\
ATST-Clip   & AS  & 79.8 \\
MW-MAE      & AS  & 80.0 \\
ATST-Frame  & AS  & 81.9 \\
Dasheng     & Mix (272k\,h) & 82.8 \\
\color{gray}PaSST       & \color{gray}AS & \color{gray}68.1 \\
\color{gray}Spatial-AST & \color{gray}AS & \color{gray}74.6 \\
\midrule
AudioSphere ($\alpha{=}0.0$)  & AS & 81.1 \\
AudioSphere ($\alpha{=}0.75$) & AS & \textbf{83.2}\,$\pm$\,0.1 \\
AudioSphere-Clean             & AS        & 83.0\,$\pm$\,0.2 \\
\bottomrule
\end{tabular}
\vspace{-8pt}
\end{wraptable}
 
 Table~\ref{tab:hear_summary} summarizes the HEAR performance of recent audio representation models and \textsc{AudioSphere} (per-task results in Appendix Table~\ref{tab:dry_audio_benchmarks}). \textsc{AudioSphere} ($\alpha{=}0.75$) reaches an aggregate of 83.2, the highest among the compared models, including Dasheng, which was trained on roughly 56 times more source audio (${\sim}$272k hours versus the ${\sim}$4.8k hours; Appendix Table~\ref{tab:baseline_summary} summarizes the pre-training data and protocol of every compared model). Dasheng's corpus contains substantially more speech, and it keeps a clear lead on speech tasks. Speech-only models (Wav2Vec 2.0, HuBERT, WavLM: 55.7--68.7) and the supervised PaSST and Spatial-AST lag in aggregate, reflecting narrower transfer.  
 
The clean data ratio study effectively reveals that the best transfer to the monaural HEAR tasks emerges from a hybrid training distribution, consistent with \citet{elsa}. This demonstrates that incorporating a proportion of clean, single-channel data during pre-training is critical for learning representations that perform well on single-channel tasks. A hybrid training distribution lifts the HEAR aggregate from 81.1 (pure simulation) to 83.2, on par with the clean-trained model ($83.0 \pm 0.2$ over three probe seeds), while simultaneously equipping the encoder with spatial sensitivity that transfers to real-world SELD (Table~\ref{tab:ablation_alpha_seld}). Crucially, the spatial pipeline matches the monaural baseline (across three downstream evaluation seeds, $83.2 \pm 0.1$ for \textsc{AudioSphere} $\alpha{=}0.75$ versus $83.0 \pm 0.2$ for \textsc{AudioSphere}-Clean) while additionally enabling frozen-probe SELD, demonstrating that spatial pre-training comes at no cost to transfer on monaural clip-level classification and frame-level event detection.

\needspace{6\baselineskip}
\subsection{\textsc{AudioSphere} Transfers to Static and Dynamic Real-World Scenes}
\label{ref:moving}

\begin{table}[t]
\centering
\caption{\textbf{Static real-world SELD on TUT-2018 REAL}. Sound event detection and direction-of-arrival estimation by number of overlapping sources. We report the mean over three downstream seeds $\pm$ the standard deviation. \textbf{Bold} = best per column within each block; ``--'' = not reported. DOAnet's LE covers only its detected frames (LR 46.5/11.5/2.9 vs.\ 83.2/48.0/26.1 for \textsc{AudioSphere}), i.e., fewer frames. $^\dagger$: frozen self-supervised backbone; other rows are supervised.}
\label{tab:static_tut}
\label{tab:static_scenes}
\small
\setlength{\tabcolsep}{5pt}
\renewcommand{\arraystretch}{1.0}
\providecommand{\pmsd}[1]{{\scriptsize\color{gray}$\pm$#1}}
\begin{tabular}{@{}l r@{\,/\,}l r@{\,/\,}l r@{\,/\,}l@{}}
\toprule
\textbf{Method} & \multicolumn{2}{c}{\textbf{Overlap 1}} & \multicolumn{2}{c}{\textbf{Overlap 2}} & \multicolumn{2}{c}{\textbf{Overlap 3}} \\
\midrule
\rowcolor{black!5}\textbf{SED} & \multicolumn{2}{c}{ER\,$\downarrow$ / F\,$\uparrow$} & \multicolumn{2}{c}{ER\,$\downarrow$ / F\,$\uparrow$} & \multicolumn{2}{c}{ER\,$\downarrow$ / F\,$\uparrow$} \\
SELDNet~\citep{seldnet} & 0.40 & 60.3 & 0.49 & 53.1 & 0.53 & 51.1 \\
MSEDnet~\citep{sednet} & 0.35 & 66.2 & \textbf{0.38} & 61.6 & \textbf{0.41} & 59.5 \\
SEDnet~\citep{sednet} & 0.38 & 64.6 & 0.42 & 61.5 & 0.43 & 57.2 \\
\addlinespace[2pt]
\textsc{AudioSphere}$^\dagger$ & \textbf{0.29}\pmsd{0.01} & \textbf{82.1}\pmsd{0.8} & \textbf{0.38}\pmsd{0.00} & \textbf{74.7}\pmsd{0.3} & 0.42\pmsd{0.01} & \textbf{71.1}\pmsd{0.2} \\
\midrule
\rowcolor{black!5}\textbf{DOA} & \multicolumn{2}{c}{LE$^\circ$\,$\downarrow$ / LR\,$\uparrow$} & \multicolumn{2}{c}{LE$^\circ$\,$\downarrow$ / LR\,$\uparrow$} & \multicolumn{2}{c}{LE$^\circ$\,$\downarrow$ / LR\,$\uparrow$} \\
SELDNet~\citep{seldnet} & 26.6 & 64.9 & 33.7 & 41.5 & 36.1 & 24.6 \\
DOAnet~\citep{doanet} & \textbf{6.3} & 46.5 & \textbf{20.1} & 11.5 & 25.8 & 2.9 \\
Spatial LibriSpeech~\citep{spatial-libri} & 12.4 & -- & \multicolumn{2}{c}{--} & \multicolumn{2}{c}{--} \\
\addlinespace[2pt]
ELSA$^\dagger$~\citep{elsa} & 15.0 & -- & \multicolumn{2}{c}{--} & \multicolumn{2}{c}{--} \\
\textsc{AudioSphere}$^\dagger$ & 13.4\pmsd{1.7} & \textbf{83.2}\pmsd{1.0} & 20.8\pmsd{1.9} & \textbf{48.0}\pmsd{1.7} & \textbf{25.3}\pmsd{2.0} & \textbf{26.1}\pmsd{2.0} \\
\bottomrule
\end{tabular}
\end{table}
 
\begin{wraptable}{r}{0.42\textwidth}
\centering
\scriptsize
\vspace{-12pt}
\setlength{\tabcolsep}{4pt}
\renewcommand{\arraystretch}{1.05}
\caption{\textbf{Static real-world SELD on TAU-2019}. Same protocol, metrics and notation as Table~\ref{tab:static_tut}. Baselines: SELDNet~\citep{seldnet}, BAT~\citep{spatialast}, w2v-SELD~\citep{w2vseld}.}
\label{tab:static_tau}
\begin{tabular}{@{}l r@{\,/\,}l r@{\,/\,}l@{}}
\toprule
& \multicolumn{2}{c}{\textbf{SED}} & \multicolumn{2}{c}{\textbf{DOA}} \\
\cmidrule(lr){2-3}\cmidrule(l){4-5}
\textbf{Method} & \multicolumn{2}{c}{ER\,$\downarrow$ / F\,$\uparrow$} & \multicolumn{2}{c}{LE$^\circ$\,$\downarrow$ / LR\,$\uparrow$} \\
\midrule
\rowcolor{black!5}\multicolumn{5}{@{}l}{\textbf{Supervised}} \\
SELDNet  & 0.34 & 79.9 & 28.5 & \textbf{85.4} \\
\rowcolor{black!5}\multicolumn{5}{@{}l}{\textbf{Self-supervised (frozen)}} \\
BAT      & 0.92 & 19.3 & 57.5 & 38.0 \\
w2v-SELD & 0.56 & 65.2 & 80.6 & 58.4 \\
\textsc{AudioSphere} & \textbf{0.28} & \textbf{84.6} & \textbf{14.1} & 78.5 \\
\hspace{1em}{\color{gray}$\pm$\,SD} & \multicolumn{2}{c}{\color{gray}0.05 / 2.1} & \multicolumn{2}{c}{\color{gray}1.3 / 0.6} \\
\bottomrule
\end{tabular}
\vspace{-8pt}
\end{wraptable}
\textbf{Static Scenes.} 
Tables~\ref{tab:static_tut} and~\ref{tab:static_tau} show that \textsc{AudioSphere} transfers to static SELD datasets TUT-2018 Real~\citep{seldnet} and TAU-2019~\citep{tau2019}. On TUT-2018 Real, the frozen model achieves the highest F-score and localization recall at every polyphony level and the lowest error rate on overlap 1. For localization error, DOAnet retains a clear lead on overlap 1 ($6.3^\circ$ vs.\ $13.4^\circ \pm 1.7^\circ$); on overlaps 2 and 3 its point estimates fall within one standard deviation of \textsc{AudioSphere}'s seed means. On TAU-2019, \textsc{AudioSphere} improves on the in-domain supervised baseline SELDNet in detection and localization while SELDNet retains higher LR, and it scores well above the reported w2v-SELD results and the evaluated BSM~+~Spatial-AST (BAT) pipeline, suggesting that patch-level self-supervision yields frame-resolved spatial features that clip-level supervised classification does not.
 
\textbf{Dynamic scenes.} Table~\ref{tab:dynamic_probing} demonstrates \textsc{AudioSphere}'s out-of-domain generalization on dynamic scenes. With a frozen backbone and shallow MLP probe, \textsc{AudioSphere} surpasses the supervised baseline on the simpler TAU-2020 (ER$_{20^{\circ}}$ 0.57 vs.\ 0.72; F$_{20^{\circ}}$ 46.6 vs.\ 37.4). However, without the temporal continuity needed to track moving sources, the MLP probe lags on the highly polyphonic TAU-2021. Attaching a lightweight temporal head (TM) to the frozen backbone recovers this performance. On TAU-2021, F$_{20^{\circ}}$ jumps from 21.0 to 37.2 and LE$_{\text{CD}}$ drops from $36.9^{\circ}$ to $24.3^{\circ}$, suggesting that \textsc{AudioSphere}'s frame-level embeddings paired with a temporal head tracks and localizes highly polyphonic sources. Furthermore, fine-tuning yields the strongest overall results both with MLP and the TM.
 
\begin{figure}[t]
\centering
\includegraphics[width=\linewidth]{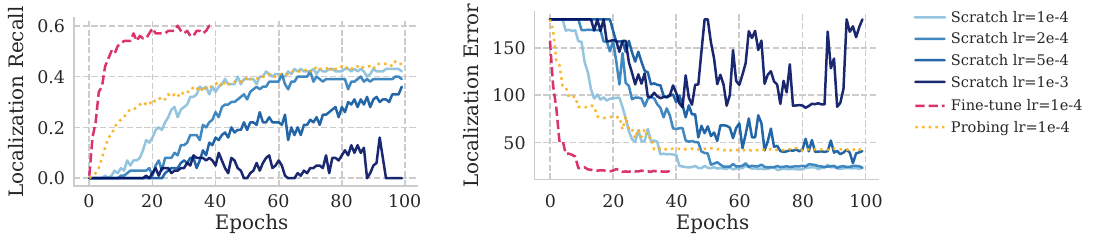}
\caption{\textbf{Training dynamics on STARSS23.} Validation localization recall and
localization error across epochs with a shallow MLP head attached to the backbone.}
\label{fig:starss23_training}
\end{figure}

\textbf{Transfer to STARSS23.} 
\autoref{fig:starss23_training} illustrates the benefit of our learned representations compared to a randomly initialized model when transferred to the STARSS23 dataset, which consists of challenging real-world dynamic recordings. Comparing validation curves confirms that \textsc{AudioSphere} provides an exceptionally strong initialization: fine-tuning converges faster, achieves lower localization error, and attains higher localization recall than training from scratch. Under frozen shallow-MLP probing, the overall localization error on STARSS23 is $48.1^{\circ}$; however, a class-level analysis (Appendix~\ref{sec:class_analysis}) reveals that this figure is heavily inflated by undetected brief sound events, which yield the maximum possible error under direct transfer. Fitting the temporal head to the frozen backbone recovers these classes, raising F$_{20^{\circ}}$ to 37.4 and reducing LE$_{\text{CD}}$ from $48.1^{\circ}$ to $22.8^{\circ}$ (Appendix~\ref{sec:class_analysis}), and additionally \autoref{fig:starss23_training} shows that fine-tuning eliminates this degradation.

\subsection{\textsc{AudioSphere} Works on Diverse Array Geometries}
\label{sec:geometry}

\begin{table}[t]
\centering
\small
\setlength{\tabcolsep}{5.5pt}
\renewcommand{\arraystretch}{1.08}
\caption{\textbf{Dynamic real-world SELD.} Frozen probing and fine-tuning on
TAU-2020/2021 with the shallow MLP probe and with the lightweight temporal head (TM)
of Section~\ref{sec:starss23}. \textbf{Bold} = best per column.}
\label{tab:dynamic_probing}
\begin{tabular}{@{}l cccc c cccc@{}}
\toprule
& \multicolumn{4}{c}{\hd{TAU-2020}} && \multicolumn{4}{c}{\hd{TAU-2021}} \\
\cmidrule(lr){2-5} \cmidrule(lr){7-10}
\hd{Method} & \hd{\ERt}\dn & \hd{\Ft}\up & \hd{\LEcd}\dn & \hd{\LRcd}\up &&
              \hd{\ERt}\dn & \hd{\Ft}\up & \hd{\LEcd}\dn & \hd{\LRcd}\up \\
\midrule
\grp{10}{Supervised baseline} \\
SELDNet          & 0.72 & 37.4 & 22.8 & 60.7 && 0.73 & 30.7 & 24.5 & 40.5 \\
\grp{10}{Probing (frozen backbone)} \\
Frozen + MLP     & 0.57 & 46.6 & 23.5 & 63.2 && 0.75 & 21.0 & 36.9 & 49.1 \\
Frozen + TM      & 0.41 & 63.6 & 15.0 & 75.0 && 0.63 & 37.2 & 24.3 & 61.3 \\
\grp{10}{Fine-tuning} \\
Fine-tune + MLP  & 0.42 & 65.8 & 13.4 & \textbf{78.2} && 0.58 & 47.9 & 18.2 & 66.3 \\
Fine-tune + TM   & \textbf{0.40} & \textbf{68.4} & \textbf{11.1} & 77.8 &&
                   \textbf{0.51} & \textbf{57.7} & \textbf{14.6} & \textbf{67.4} \\
\bottomrule
\end{tabular}
\end{table}

\begin{table}[t]
\centering
\small
\renewcommand{\arraystretch}{1.08}
\caption{\textbf{Array geometry.} (a)~Native Ambisonics vs.\ ASM-encoded tetrahedral
input on TAU2019 for SELDNet and frozen \textsc{AudioSphere}. (b)~DoA error on RSL2019-dev
(median\,$\pm$\,IQR). \textbf{Bold} = best per column.}
\label{tab:array_geometry_and_rsl2019}

\begin{subtable}[t]{0.57\linewidth}
\vspace{0pt}
\centering
\phantomsubcaption
\label{tab:array_geometry}
\setlength{\tabcolsep}{3pt}
\begin{tabular}{@{}l cccc@{}}
\toprule
\hd{Input} & \hd{ER}\dn & \hd{F}\up & \hd{LE}\dn & \hd{LR}\up \\
\midrule
\grp{5}{SELDNet (supervised)} \\
Ambisonics  & 0.34 & 79.9 & 28.5 & 85.4 \\
Mic.\ array & 0.35 & 80.0 & 30.8 & 84.0 \\
\grp{5}{\textsc{AudioSphere} (self-supervised, frozen)} \\
Ambisonics  & 0.28\pmsd{0.05} & 84.6\pmsd{2.1} & 14.1\pmsd{1.3} & 78.5\pmsd{0.6} \\
Mic.\ array & 0.25 & 84.8 & 21.2 & 73.6 \\
\bottomrule
\end{tabular}
\end{subtable}%
\hfill
\begin{subtable}[t]{0.415\linewidth}
\vspace{0pt}
\centering
\phantomsubcaption
\label{tab:rsl2019}
\setlength{\tabcolsep}{4pt}
\begin{tabular}{@{}l r@{\,$\pm$\,}l@{}}
\toprule
\hd{Method} & \multicolumn{2}{c@{}}{\hd{DoA error ($^\circ$)}\dn} \\
\midrule
\grp{3}{Baselines} \\
BSM + Spatial-AST (BAT) & 13.1 & 29.0 \\
MUSIC & 5.6 & 11.6 \\
\grp{3}{Ours} \\
ASM + \textsc{AudioSphere} & \textbf{1.9} & 6.6 \\
\bottomrule
\end{tabular}
\end{subtable}
\end{table}

\autoref{tab:array_geometry} shows that \textsc{AudioSphere} transfers to tetrahedral microphone recordings on TAU-2019 without retraining the encoder. This table demonstrates that the detection is unchanged (ER 0.28 vs.\ 0.25) and localization degrades moderately (LE $21.2^\circ$ vs.\ $14.1^\circ$). The same holds on the dynamic TAU-2020, TAU-2021 and STARSS23 datasets (Appendix~\ref{appendix:array_dynamic}, Table~\ref{tab:results_comparison}). The residual gap to native FOA is due to spatial sampling and the ASM conversion. A 4-channel tetrahedral array samples the first-order field at lower fidelity than the 32-channel Eigenmike. This observation is evidenced by SELDNet degrading by a similar amount on ASM-FOA input, and Appendix~\ref{appendix:asm_analysis} quantifying the cue degradation directly.
 
On real-world recordings of spatial sound scenes from a planar array (\autoref{tab:rsl2019}), the ASM~+~\textsc{AudioSphere} pipeline achieves the lowest median DoA error of $1.9^\circ$, compared with $5.6^\circ$ for MUSIC and $13.1^\circ$ for the evaluated BSM~+~Spatial-AST pipeline. Evaluating the frozen \textsc{AudioSphere} encoder on the same binaural renderings used for Spatial-AST (Appendix~\ref{appendix:binaural_eval}) yields a median error of $3.4^\circ$, indicating that the test-time input representation alone does not explain the difference. As front--back confusion is a common limitation of binaural input, we also report front--back folded predictions: folding reduces the Spatial-AST error from $13.1^\circ$ to about $4.3^\circ$, so front--back confusion explains a substantial but not complete part of the gap (azimuth confusion matrices in Appendix~\ref{appendix:confusion}).

\subsection{\textsc{AudioSphere} Utilizes Spatial Cues to Improve Scene Analysis}
\label{sec:res_cues}
 
We further isolate the contribution of spatial cues by constructing three reduced-input variants. In the first variant, \textsc{AudioSphere}~(W duplicated), we take the Ambisonics-pretrained model but replace the X, Y, and Z channels with copies of W at inference time, effectively stripping directional cues. In the second variant, \textsc{AudioSphere}-W is pre-trained exclusively on the omnidirectional W channel, removing all spatial information from the learned representations while retaining reverberation and noise characteristics. The last variant is \textsc{AudioSphere}-Clean, which is pre-trained solely on the AudioSet. All variants were pre-trained for 125k steps.

Table~\ref{tab:spatial_cues} (Appendix~\ref{appendix:spatial_cues}) shows that the full Ambisonics input outperforms all reduced variants across detection and localization metrics. Localization error drops from approximately $73^{\circ}$ for all three reduced variants to $14.9^{\circ}$ for the full \textsc{AudioSphere} model, confirming that inter-channel spatial structure is indispensable for accurate direction-of-arrival estimation. On the detection side, \textsc{AudioSphere}-Clean and \textsc{AudioSphere}-W perform comparably (ER 0.29 vs.\ 0.29; F 81.9 vs.\ 82.2). This suggests that the simulated reverberation and noise added during pre-training have a negligible impact on event detection, and that potential double-reverberation artifacts do not degrade performance. In contrast, \textsc{AudioSphere}~(W duplicated) yields noticeably worse detection (ER 0.33, F 79.9), likely because the model expects informative spatial variation across channels but instead receives identical copies of W, producing a distribution mismatch that degrades its learned representations. 
The full \textsc{AudioSphere} model achieves the best detection performance overall (ER 0.24, F 85.7). Altogether, these results confirm that multi-channel Ambisonics structure provides complementary information beyond what is available in the mono signal alone, benefiting both detection and localization performance.
\section{Discussion \& Conclusion}

We presented \textsc{AudioSphere}, a self-supervised spatial audio embedding model based on a multi-channel masked autoencoder that learns joint spatial-acoustic audio representations. It processes first-order Ambisonics natively and, through ASM, generalizes to diverse microphone geometries without retraining. Frozen \textsc{AudioSphere} transfers to clip-level classification of environmental sounds, speech and music, to frame-level sound event detection, and to real-world SELD. It is the first self-supervised model to acquire spatial understanding while retaining this monaural transfer. We additionally release \textsc{RealSELD}, the first probing benchmark for SELD, comprising seven real-world tasks drawn from five established datasets designed to evaluate pre-trained audio embeddings for SELD.

\textbf{Limitations.}  First, while ASM enables hardware flexibility, the quality of the Ambisonics encoding is bounded by the spatial sampling capacity of the physical hardware. For example, localization errors on TAU-2020, TAU-2021, and STARSS23 increase when replacing the 32-channel Eigenmike signal with a 4-channel tetrahedral decomposition. Native, geometry-agnostic microphone handling directly within the backbone could bypass this. Second, \textsc{AudioSphere} is pre-trained to reconstruct independent, frame-level spatial snapshots. Consequently, its frozen representations lack the intrinsic temporal-continuity priors necessary to resolve high polyphony and moving sources. A temporal probing head or fine-tuning bridges this gap, but a shallow MLP probe on a fully frozen backbone struggles to track moving mixtures. Future work could explore temporal SSL objectives, such as multi-source reconstruction over longer time windows, to learn motion priors directly from dynamic scenes. Finally, our spatial evaluation covers sound event detection and DoA estimation only; it does not evaluate distance estimation, source separation, dereverberation, or room-acoustic inference.

\textbf{Future Research.} Three avenues appear most promising. First, scaling training data using large-scale spatial corpora, such as $360^\circ$ video platforms, could reduce reliance on simulation, further narrow the sim-to-real gap, and expose the model to natural motion dynamics. Second, integrating \textsc{AudioSphere} with large language models could enable natural-language reasoning of spatial sound scenes. Third, extending \textsc{AudioSphere} to distance estimation would enable complete 3D source localization, moving beyond the current directional-only representation.

\begin{ack}
This project received funding from the NWO Talent Program (VI.Veni.202.184; KH).
This work used the Dutch national e-infrastructure with the support of the SURF
Cooperative, using grant no.\ EINF-12218. We would like to thank Robert Jan
Schlimbach from the Snellius team for helpful discussions and their help with
high-performance cluster utilization. We thank Yhonatan Gayer for the support with the SHroom framework for Ambisonics Signal Matching and for the feedback on our description of the spatial-sampling requirements of Ambisonics encoding. The authors declare no competing financial
interests. 
\end{ack}

\bibliographystyle{plainnat}
\bibliography{neurips}


\newpage
\appendix

\FloatBarrier
\section{Ambisonics}
\label{appendix:sh}

\subsection{Real Spherical Harmonics}
\label{app:sh}
The basis functions $Y_{lm}(\theta,\phi)$ are the angular part of the
solution to the wave equation in spherical coordinates, where
$\theta\in[0,2\pi)$ denotes azimuth and $\phi\in[-\pi/2,\pi/2]$ denotes
elevation. For real-valued audio signals we use real spherical
harmonics defined via the associated Legendre functions
$P_l^{|m|}(\sin\phi)$ and a normalization constant $N_{lm}$:
\begin{equation}
Y_{lm}(\theta,\phi) \;=\; N_{lm}\, P_l^{|m|}(\sin\phi)\cdot
\begin{cases}
\sin(|m|\,\theta) & m<0\\[2pt]
1                 & m=0\\[2pt]
\cos(m\,\theta)   & m>0
\end{cases}
\label{eq:sh}
\end{equation}
We adopt the three-dimensional (N3D) normalization convention used in
Ambisonics:
\begin{equation}
N_{lm}^{\mathrm{N3D}} \;=\; \sqrt{\,(2-\delta_{m,0})\,(2l+1)\,
\frac{(l-|m|)!}{(l+|m|)!}\,},
\label{eq:n3d}
\end{equation}
where $\delta_{m,0}$ is the Kronecker delta. The associated Legendre
functions are taken without the Condon--Shortley phase:
\begin{equation}
P_l^{|m|}(x) \;=\; (1-x^2)^{|m|/2}\,
\frac{d^{|m|}}{dx^{|m|}} P_l(x).
\end{equation}

\subsection{Ambisonic Channel Number (ACN) Ordering}
\label{app:acn}
We adopt the Ambisonic Channel Number (ACN) component ordering, in
which the channel index $k$ for a coefficient of order $l$ and degree
$m$ is given by
\begin{equation}
k \;=\; l(l+1)+m, \qquad 0\le l\le L,\ -l\le m\le l.
\end{equation}
For first-order Ambisonics ($L=1$), this yields the four channels
$(W,Y,Z,X)$ corresponding to $(l,m)\in\{(0,0),(1,-1),(1,0),(1,1)\}$,
with closed-form expressions
\begin{align}
W &= Y_{0,0}(\theta,\phi) = 1,\\
Y &= Y_{1,-1}(\theta,\phi) = \sqrt{3}\,\cos\phi\,\sin\theta,\\
Z &= Y_{1,0}(\theta,\phi)  = \sqrt{3}\,\sin\phi,\\
X &= Y_{1,1}(\theta,\phi)  = \sqrt{3}\,\cos\phi\,\cos\theta.
\end{align}

\FloatBarrier
\section{Ambisonics Signal Matching (ASM)}
\label{appendix:asm}

\subsection{Ambisonics Representation}
\label{appendix:ambisonics_rep}

We define the Ambisonics representation of a signal $\mathbf{s}(k)$ composed of $Q$ plane waves as
\begin{equation}
    \mathbf{a_{nm}}(k) = \mathbf{Y}_{\mathbf{\Omega}_Q}^H \mathbf{s}(k),
    \label{eq:anm_Y_Hs}
\end{equation}
where $\mathbf{Y}_{\mathbf{\Omega}_Q}$ of size $Q \times (N_a+1)^2$ is the spherical harmonics (SH) matrix. Its columns are $\mathbf{y_{nm}} = [Y_{nm}(\theta_1,\phi_1), \dots, Y_{nm}(\theta_Q,\phi_Q)]^T$, i.e. the real SH functions of order $n$ and degree $m$ evaluated at the source directions $(\theta_q,\phi_q)$.  The vector $\mathbf{a_{nm}}(k)$ collects the Ambisonics coefficients up to order $N_a$.

\subsection{Ambisonics Encoding for Arbitrary Arrays}
\label{section:Ambisonics Encoding for Arbitrary Arrays}

To bypass the geometric constraints of spherical arrays, Ambisonics encoding can be generalized to arbitrary array geometries via Tikhonov regularization.  We assume the standard spatial observation model in which the raw microphone signal vector is the sum of $Q$ plane waves and additive noise:
\begin{equation}
    \mathbf{x} = \mathbf{V}\mathbf{s} + \mathbf{n},
    \label{eq:x_Vs_n}
\end{equation}
with $\mathbf{V} \in \mathbb{C}^{M \times Q}$ the array steering matrix whose $(m,q)$-th entry is the response of microphone $m$ to a unit‑amplitude plane wave arriving from $\mathbf{\Omega}_q$, $\mathbf{s} \in \mathbb{C}^{Q \times 1}$ the plane‑wave amplitudes, and $\mathbf{n} \in \mathbb{C}^{M \times 1}$ the microphone noise vector.  Encoding is posed as a linear mapping from $\mathbf{x}$ to the estimated Ambisonics coefficients:
\begin{equation}
    \hat{a}_{nm} = \mathbf{c}_{nm}^H \mathbf{x}, \quad \forall \, 0 \leq n \leq N_a, \; -n \leq m \leq n,
\label{eq:anm_hat=Casmx}
\end{equation}
where $\mathbf{c}_{nm} \in \mathbb{C}^{M \times 1}$ is the encoding filter.  The optimal filter is obtained by minimizing the mean squared error (MSE)
\begin{equation}
    \varepsilon_{nm}^{\mathrm{ASM}} = \frac{\mathbb{E}\bigl[\lVert \hat{a}_{nm} - a_{nm} \rVert_2^2\bigr]}{\mathbb{E}\bigl[\lVert a_{nm} \rVert_2^2\bigr]}.
    \label{eq:error_nm}
\end{equation}
To obtain a signal‑independent solution we assume an isotropic diffuse sound field ($\mathbf{R}_s = \sigma_s^2\mathbf{I}$) and spatially white noise ($\mathbf{R}_n = \sigma_n^2 \mathbf{I}$).  Substituting \eqref{eq:anm_Y_Hs} and \eqref{eq:anm_hat=Casmx} into \eqref{eq:error_nm} yields
\begin{equation}
    \varepsilon_{nm}^{\mathrm{ASM}} = \frac{\sigma_s^2 \lVert\mathbf{V}^H\mathbf{c}_{nm} - \mathbf{y}_{nm}\rVert_2^2 + \sigma_n^2 \lVert\mathbf{c}_{nm}\rVert_2^2}{\sigma_s^2 \lVert\mathbf{y}_{nm}\rVert_2^2}.
\label{eq:e_nm_AMB}
\end{equation}
Minimizing \eqref{eq:e_nm_AMB} with respect to $\mathbf{c}_{nm}$ gives the optimal Ambisonics Signal Matching (ASM) filters; the term $\frac{\sigma_n^2}{\sigma_s^2}\mathbf{I}$ serves as the regularizer.  Writing the result in the row‑vector (Hermitian) form that matches \eqref{eq:anm_hat=Casmx}:
\begin{equation}
    \bigl[\mathbf{c}_{nm}^{\mathrm{ASM}}\bigr]^H = \mathbf{y}_{nm}^H \mathbf{V}^H \bigl(\mathbf{V} \mathbf{V}^H + \tfrac{\sigma_n^2}{\sigma_s^2}\mathbf{I}\bigr)^{-1}.
\label{eq:cnm=ynmV^-1}
\end{equation}
For efficient downstream application the filters can be stacked to encode all $(N_a+1)^2$ Ambisonics channels simultaneously:
\begin{equation}
    \bigl[\mathbf{C}_{\mathbf{nm}}^{\mathrm{ASM}}\bigr]^H = \bigl[\mathbf{c}_{00}^{\mathrm{ASM}}, \dots, \mathbf{c}_{N_aN_a}^{\mathrm{ASM}}\bigr]^H
    = \mathbf{Y}_{\mathbf{\Omega}_Q}^H \mathbf{V}^H \bigl(\mathbf{V} \mathbf{V}^H + \tfrac{\sigma_n^2}{\sigma_s^2}\mathbf{I}\bigr)^{-1}.
    \label{eq:Cnm=ynmV^-1}
\end{equation}

\FloatBarrier
\section{Binaural Signal Matching (BSM) for Arbitrary Arrays}
\label{appendix:bsm}

Starting from the Ambisonics encoding, binaural reproduction is obtained by
\begin{equation}
    p^{l,r}(k) = \bigl[\mathbf{h}_{\mathbf{nm}}^{l,r}(k)\bigr]^T \mathbf{a_{nm}}(k),
    \label{eq:p_hnm_anm}
\end{equation}
where $\mathbf{h}_{\mathbf{nm}}^{l,r}$ is the vector of SH coefficients of the left‑ or right‑ear HRTF expressed in the same orthonormal, Condon--Shortley‑phase basis as the Ambisonics signals.  Both the HRTF and the Ambisonics encoding are truncated to the same order $N_a = N_h$.

For nearly uniform spherical microphone arrays, plane‑wave decomposition (PWD) accurately computes these coefficients, provided the number of microphones satisfies $(N_a + 1)^2 \leq M$.

\textit{Notation: For brevity the wave number $k$ is omitted from all subsequent equations.}

Rather than routing through an intermediate Ambisonics representation, Binaural Signal Matching (BSM) extracts the binaural signals directly from the arbitrary microphone array via the linear mapping
\begin{equation}
    \hat{p}^{l,r}_{\mathrm{BSM}} = \bigl[\mathbf{c}^{l,r}_{\mathrm{BSM}}\bigr]^H \mathbf{x}.
    \label{eq:p_bsm_cx}
\end{equation}
The optimal BSM filter $\mathbf{c}^{l,r}_{\mathrm{BSM}} \in \mathbb{C}^{M \times 1}$ is obtained by minimizing the mean squared error (MSE) against the target binaural signal $p^{l,r}$:
\begin{equation}
    \varepsilon_{\mathrm{BSM}} = \mathbb{E}\bigl[\lVert \hat{p}^{l,r}_{\mathrm{BSM}} - p^{l,r} \rVert_2^2\bigr].
    \label{eq:e_bsm_mse}
\end{equation}
Substituting the observation model \eqref{eq:x_Vs_n} together with the SH representation $\mathbf{a_{nm}} = \mathbf{Y}_{\mathbf{\Omega}_Q}^H \mathbf{s}$ into \eqref{eq:p_hnm_anm} expresses the target as a linear function of the plane‑wave amplitudes, $p^{l,r} = \bigl[\mathbf{h}^{l,r}_{\mathbf{nm}}\bigr]^T \mathbf{Y}_{\mathbf{\Omega}_Q}^H \mathbf{s}$.  Applying the same isotropic diffuse field and white noise assumptions used in ASM, Tikhonov regularization yields the closed‑form BSM solution:
\begin{equation}
    \bigl[\mathbf{c}^{l,r}_{\mathrm{BSM}}\bigr]^H = \bigl[\mathbf{h}^{l,r}_{\mathbf{nm}}\bigr]^T \mathbf{Y}_{\mathbf{\Omega}_Q}^H \mathbf{V}^H \bigl( \mathbf{V} \mathbf{V}^H + \tfrac{\sigma_n^2}{\sigma_s^2}\mathbf{I} \bigr)^{-1}.
    \label{eq:c_bsm_h_v_inv}
\end{equation}

\FloatBarrier
\section{Additional Downstream Analyses}
\label{appendix:additional}

\subsection{Array-geometry transfer on dynamic scenes}
\label{appendix:array_dynamic}

Table~\ref{tab:results_comparison} extends the TAU-2019 input-format comparison of Section~\ref{sec:geometry} to the dynamic datasets. \textsc{AudioSphere} transfers to tetrahedral recordings on TAU-2020, TAU-2021, and STARSS23 without retraining the encoder, in both the frozen and fine-tuned regimes (each with the temporal head), and the fine-tuned ASM-FOA model matches or exceeds the supervised SELDNet trained directly on native Ambisonics. A residual gap to native FOA remains. The supervised SELDNet exhibits the same FOA$\to$ASM-FOA degradation when trained from scratch, confirming that the re-encoding reduces the performance also on the supervised baseline. Appendix~\ref{appendix:asm_analysis} quantifies the alignment between reference and ASM-encoded FOA channels and finds consistent azimuth and elevation cue degradation (median azimuth/elevation differences of $7.4^\circ$ / $7.6^\circ$ on TAU-2020, $8.0^\circ$ / $7.6^\circ$ on TAU-2021 and $9.4^\circ$ / $8.3^\circ$ on STARSS23).

\begin{table*}[!htb]
\centering
\caption{SELD performance on TAU2020, TAU2021, and STARSS23. \emph{FOA}: the 32-channel Eigenmike reference Ambisonics distributed with each dataset. \emph{ASM-FOA}: tetrahedral microphone signals encoded to first-order Ambisonics through our ASM pipeline. SELDNet is trained from scratch on the ASM-FOA input; \textsc{AudioSphere} is evaluated frozen and fully fine-tuned, in both cases with the temporal head (TM) of Section~\ref{sec:starss23}, i.e., the TM rows of Table ~\ref{tab:dynamic_probing}. Location-dependent metrics as in Table~\ref{tab:dynamic_probing} (Appendix~\ref{appendix:accdoa}); ER and LE: lower is better; F and LR: higher is better.}
\label{tab:results_comparison}
\resizebox{\textwidth}{!}{
\begin{tabular}{l cccc cccc cccc}
\toprule
\multirow{2}{*}{\textbf{Method}} & \multicolumn{4}{c}{\textbf{TAU2020}} & \multicolumn{4}{c}{\textbf{TAU2021}} & \multicolumn{4}{c}{\textbf{STARSS23}} \\
\cmidrule(lr){2-5} \cmidrule(lr){6-9} \cmidrule(lr){10-13}
& \ERt $\downarrow$ & \Ft $\uparrow$ & \LEcd $\downarrow$ & \LRcd $\uparrow$
& \ERt $\downarrow$ & \Ft $\uparrow$ & \LEcd $\downarrow$ & \LRcd $\uparrow$
& \ERt $\downarrow$ & \Ft $\uparrow$ & \LEcd $\downarrow$ & \LRcd $\uparrow$ \\
\midrule
\rowcolor[gray]{.95} \multicolumn{13}{l}{\textbf{Baseline (SELDNet)}} \\
Baseline FOA       & 0.72 & 37.4 & 22.8 & 60.7 & 0.73 & 30.7 & 24.5 & 40.5 & 0.57 & 29.9 & 22.0 & 47.7 \\
Baseline ASM-FOA   & 0.77 & 32.2 & 28.0 & 58.3 & 0.72 & 20.5 & 29.4 & 39.7 & 0.62 & 27.6 & 26.3 & 47.5 \\
\rowcolor[gray]{.95} \multicolumn{13}{l}{\textbf{\textsc{AudioSphere} (frozen)}} \\
Frozen + TM, FOA      & 0.41 & 63.6 & 15.0 & 75.0 & 0.63 & 37.2 & 24.3 & 61.3 & 0.56 & 37.4 & 22.8 & 58.8 \\
Frozen + TM, ASM-FOA  & 0.49 & 55.5 & 17.7 & 74.9 & 0.71 & 27.0 & 28.8 & 60.3 & 0.58 & 32.6 & 24.8 & 55.4 \\
\rowcolor[gray]{.95} \multicolumn{13}{l}{\textbf{\textsc{AudioSphere} (fine-tuned)}} \\
Fine-tune + TM, FOA     & \textbf{0.40} & \textbf{68.4} & \textbf{11.1} & \textbf{77.8} & \textbf{0.51} & \textbf{57.7} & \textbf{14.6} & \textbf{67.4} & \textbf{0.49} & \textbf{44.7} & \textbf{18.3} & \textbf{60.5} \\
Fine-tune + TM, ASM-FOA & 0.44 & 63.7 & 14.5 & 77.2 & 0.58 & 46.4 & 20.9 & 66.7 & 0.52 & 40.9 & 18.9 & 59.2 \\
\bottomrule
\end{tabular}}
\end{table*}

\subsection{Binaural-input evaluation of \textsc{AudioSphere}}
\label{appendix:binaural_eval}

To test whether the gap between the ASM~+~\textsc{AudioSphere} and BSM~+~Spatial-AST pipelines on RSL2019-dev (Table~\ref{tab:rsl2019}) can be explained by the test-time input representation alone, we evaluate the frozen \textsc{AudioSphere} encoder on the same binaural renderings used for Spatial-AST. The BSM binaural renderings are converted to first-order Ambisonics by binaural-to-Ambisonics upmixing (a Tikhonov-regularized least-squares inversion analogous to ASM). This 2-to-4-channel mapping is underdetermined and therefore close to the worst case for ASM-style conversion. Table~\ref{tab:binaural_eval} reports the median DoA error, with and without folding front--back mirrored predictions onto one hemisphere; the corresponding azimuth confusion matrices are shown in Appendix~\ref{appendix:confusion} (Figure~\ref{appendix:matrix}).

\begin{table}[!htb]
\centering
\small
\caption{Median DoA error ($^\circ$) on RSL2019-dev when both encoders receive the same binaural renderings. \emph{Folded}: predictions mirrored across the front--back axis before scoring, which removes front--back confusions.}
\label{tab:binaural_eval}
\begin{tabular}{lcc}
\toprule
\textbf{Pipeline} & \textbf{Median DoA} $\downarrow$ & \textbf{Folded} $\downarrow$ \\
\midrule
BSM + Spatial-AST (BAT) & 13.1 & \phantom{0}4.3 \\
\textsc{AudioSphere} (binaural $\rightarrow$ FOA) & \phantom{0}3.4 & \phantom{0}2.1 \\
ASM + \textsc{AudioSphere} (Table~\ref{tab:rsl2019}) & \phantom{0}1.9 & -- \\
\bottomrule
\end{tabular}
\end{table}

Operating on the same input spatial audio format as Spatial-AST, the frozen \textsc{AudioSphere} encoder attains a median DoA error of $3.4^\circ$ versus $13.1^\circ$. Folding reduces \textsc{AudioSphere}'s error from $3.4^\circ$ to $2.1^\circ$ and Spatial-AST's from $13.1^\circ$ to $4.3^\circ$: both pipelines exhibit some front--back ambiguity, but \textsc{AudioSphere} is substantially less affected, and front--back confusion accounts for a substantial but not complete part of the original gap. These results indicate that the test-time input representation alone is insufficient to explain the observed gap, but they do not identify the cause of the remaining difference, which may stem from pre-training spatialization (binaural for Spatial-AST; FOA with localized and diffuse WHAM! noise for \textsc{AudioSphere}), objective, supervision, architecture, or optimization. A native binaural \textsc{AudioSphere} variant, which would isolate encoder quality from input format, is left for future work.

\FloatBarrier
\section{Analysis of Errors Introduced by ASM}
\label{appendix:asm_analysis}

To quantify the spatial fidelity loss introduced by our ASM pipeline,
we exploit the fact that recent DCASE challenge releases ship both the
raw tetrahedral microphone signals and a reference first-order Ambisonics
(FOA) recording for every scene. We carry out this analysis on the FOA development splits of
TAU-2020, TAU-2021, and STARSS23. Because STARSS23
contains both synthetic mixtures (folds 1,2) and real recordings of
sound scenes (folds 3,4), we report these regimes
separately, giving four evaluation subsets in total.

\textbf{Methodology.}
For each paired recording we compute the broadband active intensity
vector
$\mathbf{I}(t,f) = \mathrm{Re}\bigl\{W^{*}(t,f)\,[X,Y,Z]^{\!\top}(t,f)\bigr\}$
in the short-time Fourier domain ($N_{\text{FFT}}{=}1024$, hop $256$,
$f_s{=}24$\,kHz), and average it within the $[200, 4000]$\,Hz band. The
corresponding direction of arrival (DoA) is read off as
$\theta = \mathrm{atan2}(I_y, I_x)$ (azimuth) and
$\phi = \mathrm{atan2}\!\bigl(I_z, \sqrt{I_x^{2}+I_y^{2}}\bigr)$ (elevation), following the convention of Section~\ref{sec:foa}.
We then compare the per-frame DoA estimates obtained from the reference
FOA and from the ASM-FOA, retaining only frames whose joint magnitude
(the per-frame minimum of $|W|$ across the two signals) exceeds the
50th percentile of the file. This restricts the comparison to
acoustically active content, where intensity-vector DoA is more reliable.

\begin{figure}[!htb]
    \centering
    \includegraphics[width=\linewidth]{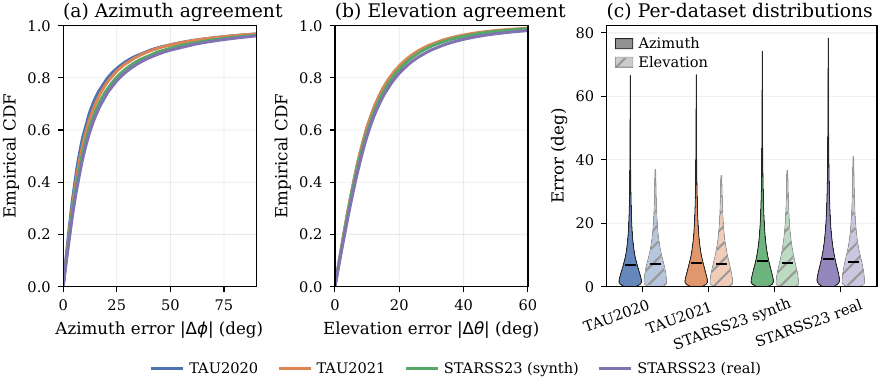}
    \caption{\textbf{Spatial-cue degradation introduced by the ASM
    pipeline.} Per-frame intensity-vector DoA agreement between the
    reference FOA and our ASM-FOA, computed in the $[200, 4000]$\,Hz band on
    energetic frames only. \textbf{(a, b)} Empirical CDFs of the
    absolute azimuth error (panel axis label $|\Delta\phi|$) and elevation error (panel axis label $|\Delta\theta|$) across the four evaluation
    subsets. \textbf{(c)} Violin plots of the same errors per dataset
    (azimuth: solid; elevation: hatched), clipped at the 95th
    percentile.}
    \label{fig:asm_doa}
\end{figure}

\textbf{Results.}
Figure~\ref{fig:asm_doa} reports the agreement between the two
intensity-vector DoA estimates as empirical CDFs of the absolute
azimuth and elevation errors, together with the per-dataset error
distributions. Median errors remain in the single-digit-degree range
across all four subsets, indicating that the spatial cues required by
downstream localization tasks are preserved by ASM. Errors are
smallest on the simulated TAU datasets, slightly larger on the
synthetic STARSS23 fold, and largest on the real STARSS23 fold.

\FloatBarrier
\section{Ablations}
\label{appendix:cleanratio}

Table~\ref{tab:ablations} reports the design ablations discussed in Section~\ref{sec:model_training}: channel-axis masking and reconstruction targets, masking pattern, masking ratio, and reconstruction loss.

\begin{table}[!htb]
\centering
\caption{Ablation results with frozen-encoder probes on ESC-50 (classification accuracy, mean\,$\pm$\,std over folds) and TAU2020 (location-dependent SELD metrics \ERt, \Ft, \LEcd, \LRcd; Appendix~\ref{appendix:accdoa}). All encoders
are pre-trained for 50k steps under an identical methodology except for the
ablated factor. 
Rows marked with $\dagger$ are trained using mask-infilling. In the mask-infilling regime the encoder sees all input tokens, but some content is replaced by learned masked tokens and extra information is added to the input to indicate which parts were masked. This allows for a fair comparison between per-channel masking, which never fully hides a token, and tube masking, which completely masks out selected tokens.
Bold marks the best value within each block.}
\label{tab:ablations}
\setlength{\tabcolsep}{4pt}
\renewcommand{\arraystretch}{1.15}
\begin{tabular}{l c c c c c}
\toprule
& \textbf{ESC-50 Acc} $\uparrow$
& \textbf{\ERt} $\downarrow$
& \textbf{\Ft} $\uparrow$
& \textbf{\LEcd} $\downarrow$
& \textbf{\LRcd} $\uparrow$ \\
\midrule
\multicolumn{6}{l}{\textit{Channel-axis masking / reconstruction target (mask-infilling regime)}} \\
Tube masking$^\dagger$          & $\mathbf{72.3 \pm 2.5}$ & $\mathbf{0.60}$ & $\mathbf{41.7}$ & $\mathbf{20.7}$ & $\mathbf{63.4}$ \\
Per-channel masking$^\dagger$   & $62.0 \pm 4.4$ & $0.66$ & $34.5$ & $22.9$ & $52.3$ \\
W$+$IVs $\rightarrow$ XYZ$^\dagger$ & $48.4 \pm 3.6$ & $0.74$ & $21.8$ & $45.4$ & $34.0$ \\
Mels $\rightarrow$ IVs$^{\dagger}$ & $33.5 \pm 2.4$ & $0.98$ & $0.8$   & $147.2$ & $3.8$ \\
\midrule
\multicolumn{6}{l}{\textit{Masking pattern (drop regime, mask ratio $0.8$)}} \\
Frequency          & $69.8 \pm 2.7$ & $0.58$ & $45.1$ & $20.1$ & $62.9$ \\
Time               & $71.2 \pm 3.2$ & $0.61$ & $40.4$ & $21.5$ & $62.4$ \\
Clustered (block)  & $73.7 \pm 1.3$ & $\mathbf{0.57}$ & $\mathbf{46.6}$ & $\mathbf{19.1}$ & $63.7$ \\
Random$^{*}$       & $\mathbf{76.1 \pm 2.2}$ & $0.58$ & $46.4$ & $19.9$ & $\mathbf{66.7}$ \\
\midrule
\multicolumn{6}{l}{\textit{Masking ratio (drop regime, random pattern)}} \\
$r=0.2$      & $71.9 \pm 4.2$ & $0.58$ & $43.0$ & $20.5$ & $63.0$ \\
$r=0.4$      & $75.0 \pm 1.7$ & $0.58$ & $45.5$ & $20.3$ & $\mathbf{67.9}$ \\
$r=0.6$      & $\mathbf{76.4 \pm 1.6}$ & $\mathbf{0.53}$ & $\mathbf{48.4}$ & $19.7$ & $67.1$ \\
$r=0.8$  & $76.1 \pm 2.2$ & $0.58$ & $46.4$ & $19.9$ & $66.7$ \\
$r=0.9$      & $74.7 \pm 1.6$ & $0.56$ & $47.5$ & $\mathbf{19.5}$ & $65.9$ \\
\midrule
\multicolumn{6}{l}{\textit{Reconstruction loss (drop regime, random pattern, $r=0.8$)}} \\
MSE (default)   & $\mathbf{76.1 \pm 2.2}$ & $\mathbf{0.58}$ & $\mathbf{46.4}$ & $19.9$ & $\mathbf{66.7}$ \\
Mel MSE $+$ IV cosine & $71.2 \pm 1.5$ & $0.60$ & $46.3$ & $\mathbf{19.2}$ & $61.6$ \\
\bottomrule
\end{tabular}
\end{table}

\begin{table}[!htb]
\centering
\small
\setlength{\tabcolsep}{4pt}
\caption{Ablation over the clean data ratio $\alpha$. Higher $\alpha$ denotes a larger share of real recordings during pre-training; $\alpha{=}0.0$ uses only simulated data on TAU-2019. Here, we evaluated the models trained for 125k steps.}
\label{tab:ablation_alpha_seld}
\begin{tabular}{l cccc}
\toprule
$\alpha$ & \textbf{ER} $\downarrow$ & \textbf{F} $\uparrow$ & \textbf{LE}$^\circ$ $\downarrow$ & \textbf{LR} $\uparrow$ \\
\midrule

$0.75$ & $\textbf{0.22} \pm 0.04$ & $\textbf{87.3} \pm 2.5$ & $14.9 \pm 0.5$ & $78.2 \pm 0.9$ \\
$0.5$  & $0.23 \pm 0.04 $ & $ 86.8\pm 2.8$ & $\textbf{14.4} \pm 0.6$ & $ \textbf{78.7} \pm 1.4$ \\
$0.25$ & $0.23 \pm 0.04$ & $ 86.8 \pm 2.5$ & $ 15.2 \pm 0.5 $ & $77.9 \pm 0.6$ \\
$0.0$  & $0.24 \pm 0.04$ & $85.7 \pm 3.1$ & $14.9 \pm 1.6$ & $ 77.0 \pm 1.5$  \\
\bottomrule
\end{tabular}
\end{table}

\textbf{Effect of clean data ratio:} We investigate the impact of the clean data ratio $\alpha$, which controls the proportion of real (non-simulated, single channel AudioSet duplicated on all Ambisonic channels) recordings relative to simulated data during pre-training. A value of $\alpha{=}0.0$ corresponds to a fully simulated training set, while higher values progressively incorporate more clean data. Table~\ref{tab:dry_audio_benchmarks} reports HEAR scores across all downstream tasks for $\alpha \in \{0.0, 0.25, 0.5, 0.75\}$.

Even a fully simulated pre-training set ($\alpha{=}0.0$) yields competitive performance, achieving an average score of $81.1$ that already surpasses most AudioSet SSL baselines except Dasheng and ATST-Frame. Progressively incorporating clean data improves the average to $83.2$ at $\alpha{=}0.75$, with the most pronounced gains in speech tasks: VL improves by $5.2$ points and CD by $4.0$ points as more real recordings are introduced. Acoustic event tasks follow a similar trend, with DCASE and ESC-50 benefiting from higher clean data ratios. Music tasks, by contrast, remain largely unaffected: Mri-S and Mri-T are saturated above $97$ across all configurations. Notably, the majority of the improvement is captured already at $\alpha{=}0.25$ ($+1.9$ points over the fully simulated baseline), indicating diminishing returns as the clean data share grows. The spike in performance after incorporating clean data signifies that \textsc{AudioSphere} with fully simulated data experiences domain mismatch, and incorporating clean data randomly serves as a remedy that improves transfer to the monaural HEAR tasks. Furthermore, TAU-2019 results suggest that incorporating clean data does not degrade the performance of \textsc{AudioSphere} on fully Ambisonics data, and even improves the performance on detection tasks. This finding is also observed in ~\citep{elsa}.

\textbf{Spatial cue supervision:} To determine whether \textsc{AudioSphere} leverages directional information for scene analysis, we evaluate on the TAU-2019 dataset under two training objectives: multi-label classification (non-spatial) and ACCDOA (spatially-aware)~\citep{shimada2021accdoa}. As shown in Table~\ref{tab:spatial_probing}, the ACCDOA objective yields a substantially lower error rate (0.28) than both the classification-only variant (0.40) and the supervised SELDNet baseline (0.34). This result indicates that the Ambisonics encoder in \textsc{AudioSphere} captures spatial structure that aids sound event discrimination, and that jointly predicting direction-of-arrival provides a useful inductive bias for high-level acoustic scene understanding.

\begin{table}[!htb]
\centering
\small
\caption{Ablation of spatial cues on TAU-2019 scene analysis. Error Rate (ER, $\downarrow$) is reported for \textsc{AudioSphere} under two training objectives and a supervised baseline.}
\label{tab:spatial_probing}
\begin{tabular}{lc}
\toprule
\textbf{Model} & \textbf{ER} $\downarrow$ \\
\midrule
\rowcolor[gray]{.95} \multicolumn{2}{l}{\textit{Supervised}} \\
SELDNet & 0.34 \\
\rowcolor[gray]{.95} \multicolumn{2}{l}{\textit{\textsc{AudioSphere}}} \\
\quad Multi-label classification & 0.40 \\
\quad ACCDOA & \textbf{0.28} \\
\bottomrule
\end{tabular}
\end{table}

\subsection{Spatial cues}
\label{appendix:spatial_cues}

\begin{table}[!htb]
    \centering
    \small
    \setlength{\tabcolsep}{6pt}
    \caption{\textbf{Ablation of spatial cues on TAU-2019.} We progressively
    remove or corrupt the spatial information available to \textsc{AudioSphere}.
    \emph{Clean} uses the audio recordings from AudioSet without reverberation and noise addition (matching to MW-MAE training), \emph{W} adds reverberation and noise back. \emph{W dup.} replaces the directional X/Y/Z
    channels with copies of W during inference. Localization metrics collapse without genuine spatial cues, confirming the model relies on directional information rather than spectral shortcuts.}
    \label{tab:spatial_cues}
    \begin{tabular}{lcccc}
    \toprule
    \textbf{Method} & \textbf{ER} $\downarrow$ & \textbf{F} $\uparrow$ & \textbf{LE}$^\circ$ $\downarrow$ & \textbf{LR} $\uparrow$ \\
    \midrule
    \textsc{AudioSphere}-Clean      & 0.29 & 81.9 & 73.0 & 64.3 \\
    \textsc{AudioSphere}-W          & 0.29 & 82.2 & 73.2 & 65.2 \\
    \textsc{AudioSphere} (W dup.)   & 0.33 & 79.9 & 73.5 & 62.5 \\
    \textsc{AudioSphere}            & \textbf{0.24} & \textbf{85.7} & \textbf{14.9} & \textbf{77.0} \\
    \bottomrule
    \end{tabular}
\end{table}

Table~\ref{tab:spatial_cues} reports the spatial-cue ablation discussed in Section~\ref{sec:res_cues}.

\FloatBarrier
\section{Downstream Evaluation}
\label{appendix:tasks}

Table~\ref{tab:tasks_overall} lists the abbreviations, task descriptions, and the types we have used to benchmark our models.

\begin{table}[!htb]
\renewcommand{\arraystretch}{1.3}
\centering
\caption{Overview of the HEAR tasks. }\resizebox{\textwidth}{!}{
\begin{tabular}{llll}
\toprule
\textbf{Abbreviation} & \textbf{Task Name} & \textbf{Description} & \textbf{Type}  \\
\midrule
DCASE & DCASE-2016 Task 2~\citep{Mesaros2018_TASLP} &  Event detection of overlapping office sounds in synthetic mixtures & Scene Analysis \\
FSD50K & FSD50k~\citep{fsd50k} &  Multilabel, large scale audio tagging & Scene Analysis \\
LC & LibriCount~\citep{libricount} & Speaker Count Identification, Simulated Cocktail Party & Scene Analysis \\
ESC-50 & ESC-50~\citep{esc50} &  Environmental Sound Classification & Environmental Sound Classification \\
CD & Crema-D~\citep{cremad} & Emotion Recognition & Speech Analysis \\
VL & VoxLingua107 Top10~\citep{voxling} & Spoken language identification & Speech Analysis \\
SC-5 & Speech Command 5h~\citep{warden2018speechcommandsdatasetlimitedvocabulary} & Keyword Spotting, reduced training subset & Speech Analysis \\
NS & NSynth Pitch 5h~\citep{nsynth2017} &  Pitch Classification, reduced training subset & Pitch Classification \\
BO & Beijing Opera~\citep{Beijingop} &  Classifying percussion instruments & Percussion \\
Mri-S &  Mridangam Stroke~\citep{mrist} &  Stroke classification in pitched percussion instruments & Percussion \\
Mri-T &  Mridangam Tonic~\citep{mrist} &  Tonic classification in pitched percussion instruments & Percussion \\
\bottomrule
\end{tabular}
}
\label{tab:tasks_overall}
\end{table}

Table~\ref{tab:dry_audio_benchmarks} reports the per-task HEAR results summarized in Table~\ref{tab:hear_summary} of the main text.

\begin{table*}[!htb]
\renewcommand{\arraystretch}{1.2}
\setlength{\tabcolsep}{3pt} 
\caption{Performance on HEAR. Reported values reflect the average performance $\pm$ standard deviation, calculated using $n$-fold cross-validation as specified by the HEAR pipeline. Bold numbers indicate the best-performing self-supervised model on the specific task. Grayed-out rows indicate supervised models. Higher $\alpha$ denotes a larger share of real recordings during pre-training; $\alpha{=}0.0$ uses only simulated data. Tasks are specified in \autoref{appendix:tasks}. \textbf{LS}: LibriSpeech 960 h, \textbf{AS}: AudioSet. The mean column is computed from the single-seed per-task values shown.}
\label{tab:dry_audio_benchmarks}
\centering
\resizebox{\textwidth}{!}{
\begin{tabular}{ll||cccc||ccc||cccc||c}
\toprule
& & \multicolumn{4}{c||}{\textbf{Acoustic Events and Scene Analysis}}  & \multicolumn{3}{c||}{\textbf{Speech}} & \multicolumn{4}{c||}{\textbf{Music}} & \multicolumn{1}{c}{}\\
\textbf{Model} & \textbf{Corpus} & \textbf{DCASE} & \textbf{FSD50K} & \textbf{LC} &  \textbf{ESC-50} & \textbf{CD} & \textbf{VL} & \textbf{SC-5}   & \textbf{NS} &  \textbf{BO} & \textbf{Mri-S} & \textbf{Mri-T} & \textbf{mean} \\
\hline

Wav2Vec 2.0 & LS & 23.5 & 29.4 & $ 69.9 \pm 2.1 $  & $ 46.4 \pm 1.8 $  & $ 57.3 \pm 1.1 $  & $ 34.9 \pm 2.4 $  & 85.3 & 17.4 & $ 81.4 \pm 4.8 $  & $ 90.7 \pm 0.8 $  & $ 77.0 \pm 0.9 $ & 55.7\\
HuBERT & LS & 78.3 & 32.8 & $ 63.3 \pm 1.2 $  & $ 58.6 \pm 2.8 $  & $ 71.2 \pm 1.2 $  & $ 65.2 \pm 2.9 $  & 94.0 & 19.8 & $ 93.2 \pm 5.9 $  & $ 94.6 \pm 0.4 $  & $ 85.0 \pm 2.5 $ & 68.7  \\
WavLM & LS & 27.0 & 25.7 & $ 61.3 \pm 2.3 $  & $ 49.5 \pm 3.8 $  & $ 64.3 \pm 1.3 $  & $ 60.1 \pm 3.2 $  & 93.8 & 18.2 & $ 84.3 \pm 6.3 $  & $ 88.8 \pm 1.0 $  & $ 76.8 \pm 0.5 $ & 59.1\\
\hline
MAE & AS & --  & 33.4 & $ 62.3 \pm 1.1 $  & $ 72.9 \pm 2.1 $  & $ 60.8 \pm 1.8 $  & $ 21.3 \pm 5.8 $  & 66.6 & 63.6 & $ 94.5 \pm 5.6 $  & $ 94.8 \pm 0.6 $  & $ 85.1 \pm 10.4 $ & 65.5\\
SSAST & AS + LS & --  & 21.4 & $ 57.8 \pm 3.3 $  & $ 58.3 \pm 2.6 $  & $ 48.0 \pm 2.1 $  & $ 15.4 \pm 2.6 $  & 22.0 & 64.2 & $ 95.8 \pm 4.3 $  & $ 90.2 \pm 5.9 $  & $ 89.1 \pm 8.0 $ & 56.2 \\
BEATs & AS & --  & 54.1 & $ 77.8 \pm 1.2 $  & $ 85.8 \pm 2.9 $  & $ 66.9 \pm 2.5 $  & $ 39.7 \pm 4.3 $  & 86.9 & 68.6 & $ 94.1 \pm 3.5 $  & $ 95.5 \pm 0.4 $  & $ 96.6 \pm 0.5 $ & 76.6 \\
MWMAE & AS & 94.2 & 51.8 & $ 80.3 \pm 1.9 $  & $ 82.2 \pm 3.2 $  & $ 74.4 \pm 1.5 $  & $ 45.5 \pm 1.7 $  & 91.6 & 69.4 & $ 95.8 \pm 4.3 $  & $ 97.5 \pm 0.4 $  & $ 97.6 \pm 0.6 $ & 80.0\\
SSAM & AS & 87.3 & 53.5 & $ 75.5 \pm 1.4 $  & $ 82.9 \pm 3.6 $  & $ 70.2 \pm 0.4 $  & $ 56.4 \pm 5.2 $  & 89.3 & 72.6 & $ 93.2 \pm 3.5 $  & $ \textbf{97.8} \pm 0.5 $  & $ 96.9 \pm 0.5 $ & 79.6  \\
Dasheng &  Mix & 94.3 & 51.2 & $75.9 \pm 2.3$ & $82.6 \pm 2.6$ & $\textbf{79.6} \pm 1.2$ & $\textbf{73.1} \pm 3.9$ & \textbf{96.6} & 71.0 & $95.3 \pm 4.6$ & $96.5 \pm 0.6$ & $94.5 \pm 0.7$ & 82.8 \\
ATST-Clip & AS & -- & \textbf{58.5} & $79.3 \pm 1.8$ & $\textbf{92.9} \pm 1.8$ & $75.9 \pm 0.6$ & $42.1 \pm 3.2$ & 93.2 & 66.8 & $95.3 \pm 3.8$ & $97.5 \pm 0.4$ & $96.2 \pm 0.7$ & 79.8 \\
ATST-Frame & AS & 94.9 & 55.9 & $75.1 \pm 2.1$ & $87.0 \pm 1.3$ & $77.5 \pm 1.3$ & $64.6 \pm 5.8$ & 91.6 & $66.4$ & $94.9 \pm 3.6$ & $97.3 \pm 0.5$ & $96.0 \pm 0.5$ & 81.9 \\
\hline
\color{gray}PASST & \color{gray}AS & \color{gray}--  & \color{gray} 64.1 & \color{gray}$ 60.7 \pm 3.7 $  & \color{gray}$ \textbf{94.8} \pm 0.3 $ & \color{gray}$ 61.8 \pm 1.1 $ & \color{gray}$ 25.9 \pm 2.6 $ & \color{gray}68.7 & \color{gray}24.2 & \color{gray}$ 96.6 \pm 3.2 $ & \color{gray}$ 96.4 \pm 0.7 $ & \color{gray}$ 87.8 \pm 1.2 $ & \color{gray}68.1 \\

\color{gray}Spatial-AST & \color{gray}AS & \color{gray}-- &  \color{gray}54.7 & \color{gray}$72.6 \pm 1.5$ & \color{gray}$90.3 \pm 1.7$ & \color{gray}$62.2 \pm 1.3$ & \color{gray}$29.1 \pm 1.9$ & \color{gray}80.6 & \color{gray}69.8 & \color{gray}$96.2 \pm 5.3$ & \color{gray}$96.2 \pm 0.4$ & \color{gray}$94.6 \pm 0.6$ & \color{gray}74.6 \\

\hline
AudioSphere ($\alpha{=}0.0$)  & AS & 94.3 & 53.0 & $79.4 \pm 1.5$ & $85.9 \pm 1.5$ & $71.9 \pm 1.9$ & $53.7 \pm 1.2$ & 89.6 & 73.8 & $94.9 \pm 4.9$ & $97.6 \pm 0.5$ & $\textbf{98.5} \pm 0.4$ & 81.1 \\
AudioSphere ($\alpha{=}0.25$) & AS & 95.1 & 54.7 & $80.8 \pm 1.7$ & $87.7 \pm 1.0$ & $75.8 \pm 1.1$ & $57.3 \pm 3.6$ & 94.5 & 75.8 & $\textbf{96.6} \pm 3.8$ & $97.5 \pm 0.5$ & $98.4 \pm 0.3$ & 83.0 \\
AudioSphere ($\alpha{=}0.5$)  & AS & 94.4 & 55.9 & $\textbf{81.7} \pm 1.6$ & $87.5 \pm 2.2$ & $75.1 \pm 1.1$ & $57.5 \pm 4.9$ & 93.0 & 75.0 & $94.9 \pm 4.9$ & $97.6 \pm 0.3$ & $98.1 \pm 0.5$ & 82.8 \\
AudioSphere ($\alpha{=}0.75$) & AS & 94.8 & 55.8 & $80.8 \pm 1.6$ & $88.3 \pm 2.1$ & $75.9 \pm 1.2$ & $58.9 \pm 3.9$ & 93.2 & \textbf{77.8} & $94.1 \pm 4.8$ & $97.6 \pm 0.3$ & $98.4 \pm 0.3$ & \textbf{83.2} \\
AudioSphere-Clean & AS & \textbf{95.3} & 56.8 & $81.3 \pm 1.8$ & $87.5 \pm 2.3$ & $75.1 \pm 0.6$ & $57.3 \pm 3.4$ & 93.5 & 75.8 & $95.8 \pm 3.7$ & $97.4 \pm 0.3$ & $98.0 \pm 0.2$ & 83.1 \\
\bottomrule
\end{tabular}
}
\end{table*}

We compare the performance and efficiency of \textsc{AudioSphere} on downstream tasks with state-of-the-art self-supervised audio representation models with a similar number of parameters as \textsc{AudioSphere} (90 M): MAE-16x16~\citep{audiomae}, SSAST-patch~\citep{gong2022ssast}, BEATs-iter3~\citep{beats}, MW-MAE-B-200-4x16~\citep{mwmae}, SSAM~\citep{audiomamba}; self-supervised speech representation models Wav2Vec 2.0 Base~\citep{wav2vec2}, HuBERT Base~\citep{hubert}, WavLM Base~\citep{wavlm}. Furthermore, we included BAT~\citep{spatialast} because it is the only model trained on spatial sound scenes. Table~\ref{tab:baseline_summary} summarizes, for each compared model, its objective and supervision, pre-training corpus and scale, native input format, evaluation-time format conversion, downstream protocol, and whether the reported numbers are reproduced by us or cited.

\begin{table}[!htb]
\centering
\caption{Baseline summary. Durations are as reported by the respective papers; the AudioSet copy used for \textsc{AudioSphere} contains ${\sim}$1.74\,M clips (${\sim}$4.8\,k nominal hours). HEAR probes use the HEAR shallow MLP protocol~\citep{hear}. SSL = self-supervised.}
\label{tab:baseline_summary}
\scriptsize
\setlength{\tabcolsep}{3pt}
\renewcommand{\arraystretch}{1.15}
\providecommand{\bsw}[1]{\dimexpr#1\dimexpr\textwidth-12\tabcolsep\relax\relax}
\begin{tabular}{@{}
  >{\raggedright\arraybackslash}p{\bsw{0.150}}
  >{\raggedright\arraybackslash}p{\bsw{0.150}}
  >{\raggedright\arraybackslash}p{\bsw{0.160}}
  >{\raggedright\arraybackslash}p{\bsw{0.105}}
  >{\raggedright\arraybackslash}p{\bsw{0.135}}
  >{\raggedright\arraybackslash}p{\bsw{0.135}}
  >{\raggedright\arraybackslash}p{\bsw{0.165}}@{}}
\toprule
\textbf{Model} & \textbf{Objective / supervision} & \textbf{Pre-training corpus (scale)} & \textbf{Native input} & \textbf{Eval-time conversion} & \textbf{Downstream protocol} & \textbf{Results} \\
\midrule
Wav2Vec 2.0, HuBERT, WavLM (Base) & SSL, masked prediction & LibriSpeech (960\,h) & mono waveform & none & HEAR MLP probe & reproduced (HEAR API) \\
MAE, SSAST, BEATs, MW-MAE, SSAM, ATST & SSL, masked autoencoding / latent prediction & AudioSet (${\sim}$4.8--5.8\,k\,h; SSAST also LibriSpeech) & mono spectrogram & none & HEAR MLP probe & reproduced (HEAR API) \\
Dasheng & SSL, masked autoencoding & Mixed (${\sim}$272\,k\,h) & mono spectrogram & none & HEAR MLP probe & reproduced (HEAR API) \\
PaSST & supervised (AudioSet labels) & AudioSet & mono spectrogram & none & HEAR MLP probe & reproduced (HEAR API) \\
\addlinespace[3pt]
Spatial-AST (BAT) & supervised (event, DoA, distance labels) & AudioSet + simulated binaural RIRs & binaural & BSM (TAU-2019, RSL2019-dev) & HEAR MLP probe (HEAR, \textsc{RealSELD}, RSL2019-dev) & reproduced \\
ELSA & contrastive (audio--text); supervised spatial-branch pre-training & Spatially augmented Clotho + AudioCaps + Freesound (${\sim}$4.7\,k\,h); Spatial LibriSpeech & FOA & -- (model not public) & -- & cited (TUT-2018 ov.\ 1 LE)$^{\dagger}$ \\
Spatial LibriSpeech & supervised (DoA) & Spatial LibriSpeech (synthetic) & FOA & -- & -- & cited (TUT-2018 ov.\ 1 LE)$^{\dagger}$ \\
w2v-SELD & SSL, wav2vec2-style & L3DAS21 + TUT-2018 + TAU-2019/20/21 (${\sim}$200\,h) & FOA waveform & none & frozen probe as reported & cited (TAU-2019) \\
\addlinespace[3pt]
SELDNet, DOAnet, MSEDnet, SEDnet & supervised, per dataset & TUT-2018 Real training split & FOA & none & -- & cited \\
SELDNet (TAU-2019/2020/2021, STARSS23) & supervised, per dataset & each dataset's training split & FOA & ASM for ASM-FOA rows & trained from scratch & cited$^{\S}$; ASM-FOA row retrained by us \\
MUSIC & signal processing, no training & -- & raw planar array & none & -- & re-implemented by us \\
\midrule
\textsc{AudioSphere} (ours) & SSL, masked reconstruction of mels + IVs & AudioSet (${\sim}$4.8\,k\,h) + simulated Ambisonics RIRs + WHAM! & FOA & ASM (tetrahedral, planar) & HEAR MLP probe; temporal head; fine-tuning & this work \\
\bottomrule
\end{tabular}
\par\smallskip
\begin{minipage}{\textwidth}\scriptsize\raggedright
$^{\S}$SELDNet numbers for TAU-2019 are from \citet{tau2019}, for TAU-2020 and STARSS23 from the official DCASE baseline repositories, and for TAU-2021 from \citet{tau2021}.
$^{\dagger}$REAL subset of TUT-2018.
\end{minipage}
\end{table}

\subsection{Extracting frame and clip level embeddings for downstream tasks} 

To obtain frame-level embeddings (80~ms resolution), we encode all patches \(P_1, \ldots, P_n\) using the frozen \textsc{AudioSphere} encoder and follow the exact patch aggregation process from~\citep{masked-modelling}. Patches are stacked and concatenated along the frequency axis, yielding one embedding per 80~ms frame that is suitable for dense prediction tasks. For clip-level embeddings, we split the input audio into non-overlapping 2-second chunks, concatenate the embedded patches of each chunk over time, and then average along the time and frequency axis. This produces a fixed-dimensional clip embedding that is independent of the original audio duration, ideal for scene classification tasks.

\FloatBarrier
\section{RealSELD}

We extend the HEAR Benchmark with ACCDOA ~\citep{shimada2021accdoa} framework to solve newly introduced real world SELD tasks. This framework jointly models sound event detection and localization across target sound classes. A class is considered active at frame $t$ when the predicted Cartesian coordinate magnitude $\|\mathbf{c}_t\| > 0.5$, where $\mathbf{c}_t \in \mathbb{R}^3$ represents the unit direction vector. We do not perform any post-processing steps unlike the HEAR protocol. All probes are the HEAR shallow MLP classifier~\citep{hear} trained on frozen embeddings; for SELD its output layer is a $3 \times C$ ACCDOA vector ($C$ classes), and the temporal head of Section~\ref{sec:starss23} replaces the MLP by a 256-d projection, frequency pooling, a GRU and multi-head self-attention in front of the same ACCDOA output.

To adapt the datasets to the HEAR format, we processed the ground truth timestamp labels into fixed-duration segments. For static sources, we extracted start and end times alongside fixed azimuth and elevation coordinates. For moving sources, we extracted active segments and their corresponding time-varying directions. In frames where no sources are active, the ACCDOA target is defined as a zero vector. Crucially, the dynamic datasets in RealSELD utilize 100ms segments (10 Hz resolution). To align the output resolution of the audio embedding model with these labels, we applied average pooling to the model representations. This ensures the \textsc{AudioSphere} model output is temporally aligned with the 100ms ground truth segments.

After the mapping, we extract the time-stamp embeddings and their corresponding ground truth labels using the HEAR evaluation kit, and train a shallow MLP probe on the extracted embeddings following the HEAR evaluation protocol. The datasets are characterized in Table~\ref{tab:RealSELD-datasets} in the main text.

\paragraph{Variance across downstream seeds}
We quantify the variance of the downstream protocol by retraining the shallow MLP probe with three seeds on the frozen backbone (Table~\ref{tab:realseld_seeds}). The same procedure on HEAR gives an aggregate of $83.2 \pm 0.1$ for \textsc{AudioSphere} ($\alpha{=}0.75$; seeds 83.2, 83.3, 83.1) and $83.0 \pm 0.2$ for \textsc{AudioSphere}-Clean (83.1, 82.8, 83.1).

\begin{table}[!htb]
\centering
\small
\caption{\textsc{RealSELD} results of the frozen \textsc{AudioSphere} backbone with a shallow MLP probe, mean $\pm$ SD over three downstream seeds. Static datasets use the location-independent ER/F and the static LE/LR; dynamic datasets use \ERt, \Ft, \LEcd, \LRcd (Appendix~\ref{appendix:accdoa}).}
\label{tab:realseld_seeds}
\setlength{\tabcolsep}{6pt}
\renewcommand{\arraystretch}{1.15}
\begin{tabular}{@{}lcccc@{}}
\toprule
\textbf{Dataset} & \textbf{ER} $\downarrow$ & \textbf{F} $\uparrow$ & \textbf{LE}$^\circ$ $\downarrow$ & \textbf{LR} $\uparrow$ \\ \midrule
\grp{5}{Static scenes (ER, F, LE, LR)} \\
TUT-2018 Real, ov.\ 1 & $0.292 \pm 0.007$ & $82.1 \pm 0.8$ & $13.41 \pm 1.71$ & $83.2 \pm 1.0$ \\
TUT-2018 Real, ov.\ 2 & $0.382 \pm 0.002$ & $74.7 \pm 0.3$ & $20.80 \pm 1.93$ & $48.0 \pm 1.7$ \\
TUT-2018 Real, ov.\ 3 & $0.423 \pm 0.006$ & $71.1 \pm 0.2$ & $25.31 \pm 2.00$ & $26.1 \pm 2.0$ \\
TAU-2019              & $0.277 \pm 0.054$ & $84.6 \pm 2.1$ & $14.05 \pm 1.26$ & $78.5 \pm 0.6$ \\
\grp{5}{Dynamic scenes (\ERt, \Ft, \LEcd, \LRcd)} \\
TAU-2020              & $0.571 \pm 0.019$ & $46.6 \pm 3.0$ & $23.54 \pm 1.86$ & $63.2 \pm 0.6$ \\
TAU-2021              & $0.747 \pm 0.007$ & $21.0 \pm 0.8$ & $36.89 \pm 0.98$ & $49.1 \pm 0.4$ \\ \bottomrule
\end{tabular}
\end{table}
\FloatBarrier
\section{Evaluation Metrics}
\label{appendix:accdoa}

We evaluate the sound localization performance on the newly generated sound localization tasks, and RSL2019-dev by calculating the Direction of Arrival (DoA) error $\theta$ between the [$x,y,z$] coordinates of the target sound source on the unit sphere using the arc cosine of the dot product of the unit vectors: $\theta = \arccos(v \cdot  \hat{v})$.

\textbf{SELD metrics on RealSELD Dynamic Motion: } We used four joint localization and detection metrics ~\citep{eval_overview}, which are widely-used in audio-only SELD tasks ~\citep{STARSS22}. Two metrics, referred to as location-aware detection, are the error rate (ER$_{20^{\circ}}$) and F-score (F$_{20^{\circ}}$) for one-second non-overlapping segments. We consider a prediction to be a true positive (TP) if the prediction and the reference class are the same and the angle difference is less than $20^{\circ}$. F$_{20^{\circ}}$ is calculated from location-aware precision and recall, whereas ER$_{20^{\circ}}$ is the sum of insertion, deletion, and substitution errors, divided by the total number of references. The other two metrics, referred to as class-aware localization, are localization error (LE$_{\text{CD}}$) in degrees and localization recall (LR$_{\text{CD}}$) in one-second non-overlapping segments, where the subscript denotes classification-dependent. Unlike location-aware detection, we do not use a threshold; instead, we estimate the difference between the correct prediction and the reference. LE$_{CD}$ expresses the average angular difference between the same class’s predictions and references. LR$_{CD}$ reports the true positive rate as the number of localization estimates detected in a class out of the total number of class instances. We used the macro mode of computation, which does not apply to ER$_{20^{\circ}}$ because it includes substitution errors between two classes. We first computed the metrics for each class and then averaged them for the other three metrics to obtain the final system performance. 

\textbf{SELD metrics on RealSELD Static Motion: }For testing the model on the TUT-2018 REAL and TAU-2019, we used the non-localization-dependent counterparts of the aforementioned metrics. ER and F-score calculated in segments of one second with no overlap, as proposed. The segment-wise results are obtained from the classifier's frame-level predictions, treating a sound event as active across the entire segment if it is active in any frame. Similarly, we obtain labels for one-second segments of the reference from its framewise annotation and calculate segment-wise ER and F-scores. Additionally, in order to account for time frames where the number of estimated and reference DOAs are unequal, we report the localization recall, calculated as ${\text{TP}}/{\text{(TP + FN)}}$ in percentage, where true positives TP is the total number of time frames in which the number of DOAs predicted is equal to reference, and false negatives FN is the total number of frames where the predicted and reference DOA are unequal.

\FloatBarrier
\section{Class-Level Analysis for STARSS23}
\label{sec:class_analysis}

To better understand the behavior of \textsc{AudioSphere} with a shallow MLP probe during out-of-domain evaluation, we provide a granular breakdown of its performance across the 13 target classes of the STARSS23 dataset. Table~\ref{tab:class_wise} details the location-dependent F-score ($F_{20^\circ}$) and the localization error ($LE_{CD}$) for the direct transfer setting.

\begin{table}[!htb]
\centering
\small
\caption{Class-wise metrics for STARSS23 Direct Transfer. We report location-dependent F-score ($F_{20^\circ}$, higher is better) and Localization Error ($LE_{CD}$, lower is better). The DCASE convention assigns maximum error ($180^{\circ}$) to undetected classes.}
\label{tab:class_wise}
\setlength{\tabcolsep}{10pt} 
\renewcommand{\arraystretch}{1.15} 
\begin{tabular}{@{}lcc@{}}
\toprule
\textbf{Class} & \boldmath{$F_{20^\circ}$} $\uparrow$ & \boldmath{$LE_{CD}$} $\downarrow$ \\ \midrule
Fem. Speech     & 0.52   & 25.27 \\
Male Speech     & 0.63   & 15.17 \\
Clap            & 0.20   & 18.19 \\
Phone           & 0.01   & 36.21 \\
Laugh           & 0.36   & 17.63 \\
Domestic sounds & 0.30   & 20.54 \\
Footsteps       & 0.01   & 37.45 \\
Door            & 0.00   & 180.00 \\
Music           & 0.24   & 37.12 \\
Music Instr.    & 0.40   & 20.91 \\
Faucet          & 0.28   & 13.19 \\
Bell            & 0.00   & 180.00 \\
Knock           & 0.13   & 10.31 \\ \bottomrule
\end{tabular}
\end{table} 

\paragraph{Strong Performance on Speech and Harmonic Sounds} 
\textsc{AudioSphere} demonstrates exceptional direct-transfer capabilities on human vocalizations. ``Male Speech'' and ``Female Speech'' are among the highest performing categories, achieving $F_{20^\circ}$ scores of 0.63 and 0.52, respectively, alongside strong localization precision. Similarly, continuous and harmonic sounds such as ``Laugh'' ($17.63^\circ$) and ``Music Instr.'' ($20.91^\circ$) yield highly robust localization. This suggests that our spatial-acoustic pre-training excels at learning stable representations of sounds with sustained spectral energy and frequent occurrence across diverse acoustic environments.

\paragraph{Failure Modes and Limitations} 
Conversely, the model struggles significantly with specific environmental transients and quiet events, notably ``Door,'' ``Bell,'' and ``Footsteps.'' For ``Door'' and ``Bell,'' the model fails to successfully detect and localize the events within the $20^\circ$ tolerance threshold, resulting in $F_{20^\circ}$ scores of 0.00 and a maximum localization error penalty of $180.00^\circ$. Footsteps ($F_{20^\circ}$ of 0.01) similarly suffer, likely because they are low-energy events heavily susceptible to masking by reverberation and ambient noise in real-world dynamic scenes. These failure cases highlight an ongoing challenge in zero-shot SELD transfer: bridging the gap for rare, highly reverberant, or low-energy impulsive sounds.

\paragraph{Effect of temporal integration}
We hypothesized that limited temporal integration is a main contributor to the failures on brief, intermittent, low-energy or moving events. Fitting the SELDNet-style temporal head of Section~\ref{sec:starss23} to the frozen backbone (overall STARSS23: F$_{20^\circ}$ 37.4, LE$_{CD}$ $22.8^\circ$, LR$_{CD}$ 58.8) improves these classes (Table~\ref{tab:class_wise_tm}) and thereby reduces class-wise variability. This experiment demonstrates that the frozen frame-level embeddings carry spatial-acoustic information that a standard sequence model can integrate over time.

\begin{table}[!htb]
\centering
\small
\caption{STARSS23 failure classes with the shallow MLP probe versus the temporal head on the same frozen backbone.}
\label{tab:class_wise_tm}
\setlength{\tabcolsep}{8pt}
\renewcommand{\arraystretch}{1.15}
\begin{tabular}{@{}lcc@{}}
\toprule
\textbf{Class} & \boldmath{$F_{20^\circ}$} $\uparrow$ (MLP $\rightarrow$ temporal) & \boldmath{$LE_{CD}$} $\downarrow$ (MLP $\rightarrow$ temporal) \\ \midrule
Door      & 0.00 $\rightarrow$ 0.30 & 180.0 $\rightarrow$ 11.4 \\
Bell      & 0.00 $\rightarrow$ 0.17 & 180.0 $\rightarrow$ 31.7 \\
Footsteps & 0.01 $\rightarrow$ 0.12 & \phantom{0}37.5 $\rightarrow$ 27.1 \\
Phone     & 0.01 $\rightarrow$ 0.16 & \phantom{0}36.2 $\rightarrow$ 26.8 \\ \bottomrule
\end{tabular}
\end{table}

\FloatBarrier
\section{SoundSpaces 2.0 Specifications}
\label{appendix:brir-spec}

We generate our ARIRs using the simulator provided by SoundSpaces 2.0~\citep{chen22soundspaces2}. Our hyperparameters for the simulator is depicted in Table \ref{tab:acoustic_config}.

\begin{table}[!htb]
\caption{Acoustic configuration parameters utilized in SoundSpaces 2.0 to generate our ARIRs.}
\label{tab:acoustic_config}
\centering
\small
\begin{tabular}{lcllc}
\toprule
\textbf{Parameter} & \textbf{Value} & \phantom{xxxx} & \textbf{Parameter} & \textbf{Value} \\
\midrule
directSHOrder & 3 && indirectSHOrder & 3 \\
sampleRate & 32000 && frequencyBands & 8 \\
maxDiffractionOrder & 10 && transmission & True \\
indirect & True && indirectRayCount & 15000 \\
indirectRayDepth & 400 && sourceRayCount & 200 \\
sourceRayDepth & 20 && threadCount & 16 \\
agentHeight & 1.5m \\
\bottomrule
\end{tabular}
\end{table}

\FloatBarrier
\section{Detailed Pre-Training HyperParameters}

Table~\ref{tab:hyperparams} summarizes all hyperparameters. We train \textsc{AudioSphere} for 500K steps on a single H100 96\,GB GPU with 16 CPU cores, minimizing the mean squared error (MSE) between predicted and ground-truth patches at masked positions. We use AdamW~\citep{loshchilov2017decoupled} with weight decay $0.01$, gradient clipping at max norm $5.0$, and a cosine learning rate schedule with 10K-step warm-up from $2 \times 10^{-4}$ decaying to $0$. Training converges in approximately 90 GPU hours.

\begin{table}[!htb]
\centering
\caption{Pre-training hyperparameters for \textsc{AudioSphere}.}
\label{tab:hyperparams}
\small
\begin{tabular}{ll}
\toprule
\textbf{Parameter} & \textbf{Value} \\
\midrule
Encoder & ViT-Base (12 layers, 768 dim, 12 heads) \\
Decoder & 8 layers, local-global attention \\
Patch size & $7 \times 8 \times 16$ (channels $\times$ time $\times$ freq) \\
Embedding dim & 768 \\
Mask ratio & 80\% (tube masking) \\
Segment length & 2\,s (200 frames) \\
Segments per clip & 16 \\
Nominal / effective batch size & 96 / 1{,}536 \\
Optimizer & AdamW ($\beta_1{=}0.9$, $\beta_2{=}0.95$) \\
Learning rate & $2 \times 10^{-4}$ (cosine decay to 0) \\
Warm-up steps & 10{,}000 \\
Weight decay & 0.01 \\
Gradient clipping & max norm 5.0 \\
Training steps & 500K \\
\bottomrule
\end{tabular}
\end{table}

\FloatBarrier
\clearpage
\section{Confusion Matrices for BAT and \textsc{AudioSphere} on RSL2019-dev}
\label{appendix:confusion}

\begin{figure}[!htb]
    \centering
    \includegraphics[width=\linewidth]{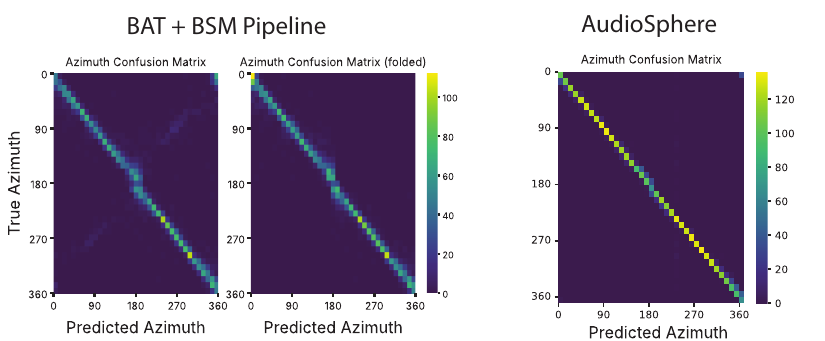}
    \caption{\textbf{Confusion Matrices for BAT + BSM Pipeline and \textsc{AudioSphere}.} Here, we depict the azimuth confusion matrices for BAT and \textsc{AudioSphere} on RSL2019-dev dataset. BAT + BSM's errors concentrate on front--back mirrored directions; see Appendix~\ref{appendix:binaural_eval} for the folded analysis.}
    \label{appendix:matrix}
\end{figure}


\newpage
\input{checklist.tex}

\end{document}

%% file: checklist.tex
\section*{NeurIPS Paper Checklist}

\begin{enumerate}

\item {\bf Claims}
    \item[] Question: Do the main claims made in the abstract and introduction accurately reflect the paper's contributions and scope?
    \item[] Answer: \answerYes{} 
    \item[] Justification: Our main claims are: (i) \textsc{AudioSphere} is a self-supervised spatial-acoustic representation model whose frozen embeddings transfer, with a shallow MLP probe, to clip-level classification of environmental sound, speech and music and to frame-level sound event detection (HEAR, Section 5.1), and to real-world sound event localization and detection (RealSELD, Section 5.2). (ii) RealSELD, an evaluation kit that adapts five real-world SELD datasets (seven tasks) to a frozen-backbone protocol (Section 4, Appendix H). (iii) Microphone flexibility via ASM, supported by the ASM literature and by our experiments on tetrahedral arrays (TAU-2019 in Section 5.3; TAU-2020, TAU-2021 and STARSS23 in Appendix D) and on a planar array (RSL2019-dev). The comparison with Spatial-AST on RSL2019-dev is stated at the level of complete evaluated pipelines (Section 5.3, Appendix D), and the scope of the spatial evaluation (detection and direction-of-arrival estimation only) is stated in Section 6.
 
\item {\bf Limitations}
    \item[] Question: Does the paper discuss the limitations of the work performed by the authors?
    \item[] Answer: \answerYes{}{} 
    \item[] Justification: In Section 6 we discuss the limitations of our work.
 
\item {\bf Theory assumptions and proofs}
    \item[] Question: For each theoretical result, does the paper provide the full set of assumptions and a complete (and correct) proof?
    \item[] Answer: \answerNA{}.
    \item[] Justification: Our paper does not derive new theoretical results.
 
    \item {\bf Experimental result reproducibility}
    \item[] Question: Does the paper fully disclose all the information needed to reproduce the main experimental results of the paper to the extent that it affects the main claims and/or conclusions of the paper (regardless of whether the code and data are provided or not)?
    \item[] Answer: \answerYes{} 
    \item[] Justification: We describe the data-generation pipeline, model and training objective in Section 3 and the evaluation protocols in Section 4, with full details in the appendices: ASM encoding (B), the HEAR tasks, embedding extraction and the protocol of every baseline (G), the construction of RealSELD and the downstream heads (H), evaluation metrics (I), simulator settings (K) and pre-training hyperparameters (L).

\item {\bf Open access to data and code}
    \item[] Question: Does the paper provide open access to the data and code, with sufficient instructions to faithfully reproduce the main experimental results, as described in supplemental material?
    \item[] Answer: \answerYes{} 
    \item[] Justification: Code, pre-trained checkpoints, the RealSELD evaluation kit (scripts and dataset-split files) and the preprocessing pipeline are released on GitHub and HuggingFace (footnote 1).

\item {\bf Experimental setting/details}
    \item[] Question: Does the paper specify all the training and test details (e.g., data splits, hyperparameters, how they were chosen, type of optimizer) necessary to understand the results?
    \item[] Answer: \answerYes{} 
    \item[] Justification: Pre-training data generation is described in Section 3 and Appendix K, and the optimizer and all training hyperparameters in Appendix L. The downstream protocols (shallow MLP probing following HEAR, the temporal head, fine-tuning and its learning-rate sweep) are given in Section 4 and Appendix H, and the RSL2019-dev splits in Section 4.
 
\item {\bf Experiment statistical significance}
    \item[] Question: Does the paper report error bars suitably and correctly defined or other appropriate information about the statistical significance of the experiments?
    \item[] Answer: \answerYes{} 
    \item[] Justification: Error bars capture two sources of variability: (i) $n$-fold cross-validation as specified by the HEAR pipeline, mean $\pm$ standard deviation (per-task HEAR results, Appendix G); and (ii) variability over three downstream seeds of the probe trained on the frozen backbone, mean $\pm$ one standard deviation (HEAR aggregate in Section 5.1; RealSELD static and dynamic results in Section 5.2 and Appendix H).
 
\item {\bf Experiments compute resources}
    \item[] Question: For each experiment, does the paper provide sufficient information on the computer resources (type of compute workers, memory, time of execution) needed to reproduce the experiments?
    \item[] Answer: \answerYes{} 
    \item[] Justification: Appendix L reports the hardware (a single H100 96\,GB GPU with 16 CPU cores) and the pre-training cost (${\sim}$90 GPU hours for 500k steps). Ablation models were pre-trained for 50k steps (Appendix F) or 125k steps (Section 5.4), and downstream probes were repeated over three seeds (Appendix H).
    
\item {\bf Code of ethics}
    \item[] Question: Does the research conducted in the paper conform, in every respect, with the NeurIPS Code of Ethics \url{https://neurips.cc/public/EthicsGuidelines}?
    \item[] Answer: \answerYes{} 
    \item[] Justification: Our paper conforms with the ethics.

\item {\bf Broader impacts}
    \item[] Question: Does the paper discuss both potential positive societal impacts and negative societal impacts of the work performed?
    \item[] Answer: \answerNA{} 
    \item[] Justification: Our work addresses spatial audio representation learning, and does not possess negative societal impacts.
    
\item {\bf Safeguards}
    \item[] Question: Does the paper describe safeguards that have been put in place for responsible release of data or models that have a high risk for misuse (e.g., pre-trained language models, image generators, or scraped datasets)?
    \item[] Answer: \answerNA{} 
    \item[] Justification: Our work addresses spatial audio representation learning, and does not possess negative societal impacts.
 
\item {\bf Licenses for existing assets}
    \item[] Question: Are the creators or original owners of assets (e.g., code, data, models), used in the paper, properly credited and are the license and terms of use explicitly mentioned and properly respected?
    \item[] Answer: \answerYes{}
    \item[] Justification: We cited each dataset and the model code we've used. Furthermore, our released datasets and benchmarks include all necessary licenses.
 
\item {\bf New assets}
    \item[] Question: Are new assets introduced in the paper well documented and is the documentation provided alongside the assets?
    \item[] Answer: \answerYes{} 
    \item[] Justification: The \textsc{AudioSphere} checkpoints and pre-training data are released on HuggingFace and the RealSELD evaluation kit on GitHub, each documented alongside the asset.
 
\item {\bf Crowdsourcing and research with human subjects}
    \item[] Question: For crowdsourcing experiments and research with human subjects, does the paper include the full text of instructions given to participants and screenshots, if applicable, as well as details about compensation (if any)? 
    \item[] Answer: \answerNA{}{} 
    \item[] Justification: We did not perform crowdsourcing or research with human subjects.
 
\item {\bf Institutional review board (IRB) approvals or equivalent for research with human subjects}
    \item[] Question: Does the paper describe potential risks incurred by study participants, whether such risks were disclosed to the subjects, and whether Institutional Review Board (IRB) approvals (or an equivalent approval/review based on the requirements of your country or institution) were obtained?
    \item[] Answer: \answerNA{} 
    \item[] Justification: Our paper does not involve crowdsourcing nor research with human subjects
 
\item {\bf Declaration of LLM usage}
    \item[] Question: Does the paper describe the usage of LLMs if it is an important, original, or non-standard component of the core methods in this research? Note that if the LLM is used only for writing, editing, or formatting purposes and does \emph{not} impact the core methodology, scientific rigor, or originality of the research, declaration is not required.
    \item[] Answer: \answerNA{} 
    \item[] Justification: We did not use LLMs to develop methodological parts of this research.

\end{enumerate}

%% file: neurips_2026.bbl
\begin{thebibliography}{65}
\providecommand{\natexlab}[1]{#1}
\providecommand{\url}[1]{\texttt{#1}}
\expandafter\ifx\csname urlstyle\endcsname\relax
  \providecommand{\doi}[1]{doi: #1}\else
  \providecommand{\doi}{doi: \begingroup \urlstyle{rm}\Url}\fi

\bibitem[Adavanne et~al.(2017)Adavanne, Pertilä, and Virtanen]{sednet}
Sharath Adavanne, Pasi Pertilä, and Tuomas Virtanen.
\newblock Sound event detection using spatial features and convolutional recurrent neural network.
\newblock In \emph{2017 IEEE International Conference on Acoustics, Speech and Signal Processing (ICASSP)}, pages 771--775, 2017.
\newblock \doi{10.1109/ICASSP.2017.7952260}.

\bibitem[Adavanne et~al.(2018)Adavanne, Politis, and Virtanen]{doanet}
Sharath Adavanne, Archontis Politis, and Tuomas Virtanen.
\newblock Direction of arrival estimation for multiple sound sources using convolutional recurrent neural network.
\newblock In \emph{2018 26th European Signal Processing Conference (EUSIPCO)}, pages 1462--1466, 2018.
\newblock \doi{10.23919/EUSIPCO.2018.8553182}.

\bibitem[Adavanne et~al.(2019{\natexlab{a}})Adavanne, Politis, Nikunen, and Virtanen]{seldnet}
Sharath Adavanne, Archontis Politis, Joonas Nikunen, and Tuomas Virtanen.
\newblock Sound event localization and detection of overlapping sources using convolutional recurrent neural networks.
\newblock \emph{IEEE Journal of Selected Topics in Signal Processing}, 13\penalty0 (1):\penalty0 34--48, 2019{\natexlab{a}}.
\newblock \doi{10.1109/JSTSP.2018.2885636}.

\bibitem[Adavanne et~al.(2019{\natexlab{b}})Adavanne, Politis, and Virtanen]{tau2019}
Sharath Adavanne, Archontis Politis, and Tuomas Virtanen.
\newblock A multi-room reverberant dataset for sound event localization and detection.
\newblock In \emph{Proceedings of the Detection and Classification of Acoustic Scenes and Events 2019 Workshop ({DCASE2019})}, pages 10--14, New York University, NY, USA, October 2019{\natexlab{b}}.
\newblock URL \url{https://dcase.community/workshop2019/proceedings}.

\bibitem[Anantapadmanabhan et~al.(2013)Anantapadmanabhan, Bellur, and Murthy]{mrist}
Akshay Anantapadmanabhan, Ashwin Bellur, and Hema~A Murthy.
\newblock Modal analysis and transcription of strokes of the mridangam using non-negative matrix factorization.
\newblock In \emph{2013 IEEE International Conference on Acoustics, Speech and Signal Processing}, pages 181--185, 2013.
\newblock \doi{10.1109/ICASSP.2013.6637633}.

\bibitem[Baade et~al.(2022)Baade, Peng, and Harwath]{mae-ast}
Alan Baade, Puyuan Peng, and David Harwath.
\newblock {MAE-AST}: Masked autoencoding audio spectrogram transformer.
\newblock In \emph{Interspeech 2022}, pages 2438--2442, 2022.
\newblock \doi{10.21437/Interspeech.2022-10961}.

\bibitem[Baevski et~al.(2020)Baevski, Zhou, Mohamed, and Auli]{wav2vec2}
Alexei Baevski, Henry Zhou, Abdelrahman Mohamed, and Michael Auli.
\newblock wav2vec 2.0: a framework for self-supervised learning of speech representations.
\newblock In \emph{Proceedings of the 34th International Conference on Neural Information Processing Systems}, NIPS '20, Red Hook, NY, USA, 2020. Curran Associates Inc.
\newblock ISBN 9781713829546.

\bibitem[Bastine et~al.(2022)Bastine, Birnie, Abhayapala, Samarasinghe, and Tourbabin]{ref_ben}
Amy Bastine, Lachlan Birnie, Thushara~D. Abhayapala, Prasanga Samarasinghe, and Vladimir Tourbabin.
\newblock Ambisonics capture using microphones on head-worn device of arbitrary geometry.
\newblock In \emph{2022 30th European Signal Processing Conference (EUSIPCO)}, pages 309--313, 2022.
\newblock \doi{10.23919/EUSIPCO55093.2022.9909803}.

\bibitem[Bregman(1984)]{bregman1984auditory}
Albert~S Bregman.
\newblock Auditory scene analysis.
\newblock In \emph{Proceedings of the 7th International Conference on Pattern Recognition}, pages 168--175. Citeseer, 1984.

\bibitem[Cao et~al.(2014)Cao, Cooper, Keutmann, Gur, Nenkova, and Verma]{cremad}
Houwei Cao, David~G Cooper, Michael~K Keutmann, Ruben~C Gur, Ani Nenkova, and Ragini Verma.
\newblock {CREMA-D}: Crowd-sourced emotional multimodal actors dataset.
\newblock \emph{IEEE Trans Affect Comput}, 5\penalty0 (4):\penalty0 377--390, October 2014.

\bibitem[Chang et~al.(2017)Chang, Dai, Funkhouser, Halber, Niessner, Savva, Song, Zeng, and Zhang]{Matterport3D}
Angel Chang, Angela Dai, Thomas Funkhouser, Maciej Halber, Matthias Niessner, Manolis Savva, Shuran Song, Andy Zeng, and Yinda Zhang.
\newblock {Matterport3D}: Learning from {RGB-D} data in indoor environments.
\newblock \emph{International Conference on 3D Vision (3DV)}, 2017.

\bibitem[Chen et~al.(2022)Chen, Schissler, Garg, Kobernik, Clegg, Calamia, Batra, Robinson, and Grauman]{chen22soundspaces2}
Changan Chen, Carl Schissler, Sanchit Garg, Philip Kobernik, Alexander Clegg, Paul Calamia, Dhruv Batra, Philip~W Robinson, and Kristen Grauman.
\newblock Soundspaces 2.0: A simulation platform for visual-acoustic learning.
\newblock In \emph{NeurIPS 2022 Datasets and Benchmarks Track}, 2022.

\bibitem[Chen et~al.(2021)Chen, Wang, Chen, Wu, Liu, Chen, Li, Kanda, Yoshioka, Xiao, Wu, Zhou, Ren, Qian, Qian, Wu, Zeng, and Wei]{wavlm}
Sanyuan Chen, Chengyi Wang, Zhengyang Chen, Yu~Wu, Shujie Liu, Zhuo Chen, Jinyu Li, Naoyuki Kanda, Takuya Yoshioka, Xiong Xiao, Jian Wu, Long Zhou, Shuo Ren, Yanmin Qian, Yao Qian, Jian Wu, Michael Zeng, and Furu Wei.
\newblock {WavLM}: Large-scale self-supervised pre-training for full stack speech processing.
\newblock \emph{CoRR}, abs/2110.13900, 2021.
\newblock URL \url{http://dblp.uni-trier.de/db/journals/corr/corr2110.html#abs-2110-13900}.

\bibitem[Chen et~al.(2023)Chen, Wu, Wang, Liu, Tompkins, Chen, Che, Yu, and Wei]{beats}
Sanyuan Chen, Yu~Wu, Chengyi Wang, Shujie Liu, Daniel Tompkins, Zhuo Chen, Wanxiang Che, Xiangzhan Yu, and Furu Wei.
\newblock {BEAT}s: Audio pre-training with acoustic tokenizers.
\newblock In Andreas Krause, Emma Brunskill, Kyunghyun Cho, Barbara Engelhardt, Sivan Sabato, and Jonathan Scarlett, editors, \emph{Proceedings of the 40th International Conference on Machine Learning}, volume 202 of \emph{Proceedings of Machine Learning Research}, pages 5178--5193. PMLR, 23--29 Jul 2023.
\newblock URL \url{https://proceedings.mlr.press/v202/chen23ag.html}.

\bibitem[Chong et~al.(2023)Chong, Wang, Zhou, and Zeng]{MaskSpec}
Dading Chong, Helin Wang, Peilin Zhou, and Qingcheng Zeng.
\newblock Masked spectrogram prediction for self-supervised audio pre-training.
\newblock In \emph{ICASSP 2023 - 2023 IEEE International Conference on Acoustics, Speech and Signal Processing (ICASSP)}, pages 1--5, 2023.
\newblock \doi{10.1109/ICASSP49357.2023.10095691}.

\bibitem[Devnani et~al.(2024)Devnani, Seto, Aldeneh, Toso, Menyaylenko, Theobald, Sheaffer, and Sarabia]{elsa}
Bhavika Devnani, Skyler Seto, Zakaria Aldeneh, Alessandro Toso, Elena Menyaylenko, Barry-John Theobald, Jonathan Sheaffer, and Miguel Sarabia.
\newblock Learning spatially-aware language and audio embeddings.
\newblock In A.~Globerson, L.~Mackey, D.~Belgrave, A.~Fan, U.~Paquet, J.~Tomczak, and C.~Zhang, editors, \emph{Advances in Neural Information Processing Systems}, volume~37, pages 33505--33537. Curran Associates, Inc., 2024.
\newblock \doi{10.52202/079017-1056}.

\bibitem[Dinkel et~al.(2024)Dinkel, Yan, Wang, Zhang, Wang, and Wang]{dasheng}
Heinrich Dinkel, Zhiyong Yan, Yongqing Wang, Junbo Zhang, Yujun Wang, and Bin Wang.
\newblock Scaling up masked audio encoder learning for general audio classification.
\newblock In \emph{Interspeech 2024}, 2024.

\bibitem[dos Santos et~al.(2023)dos Santos, Rosero, and de~Alencar~Lotufo]{w2vseld}
Orlem~Lima dos Santos, Karen Rosero, and Roberto de~Alencar~Lotufo.
\newblock {w2v-SELD}: A sound event localization and detection framework for self-supervised spatial audio pre-training, 2023.
\newblock URL \url{https://arxiv.org/abs/2312.06907}.

\bibitem[Dosovitskiy et~al.(2021)Dosovitskiy, Beyer, Kolesnikov, Weissenborn, Zhai, Unterthiner, Dehghani, Minderer, Heigold, Gelly, Uszkoreit, and Houlsby]{dosovitskiy2020vit}
Alexey Dosovitskiy, Lucas Beyer, Alexander Kolesnikov, Dirk Weissenborn, Xiaohua Zhai, Thomas Unterthiner, Mostafa Dehghani, Matthias Minderer, Georg Heigold, Sylvain Gelly, Jakob Uszkoreit, and Neil Houlsby.
\newblock An image is worth 16x16 words: Transformers for image recognition at scale.
\newblock \emph{ICLR}, 2021.

\bibitem[Engel et~al.(2017)Engel, Resnick, Roberts, Dieleman, Eck, Simonyan, and Norouzi]{nsynth2017}
Jesse Engel, Cinjon Resnick, Adam Roberts, Sander Dieleman, Douglas Eck, Karen Simonyan, and Mohammad Norouzi.
\newblock Neural audio synthesis of musical notes with wavenet autoencoders, 2017.

\bibitem[Fonseca et~al.(2022)Fonseca, Favory, Pons, Font, and Serra]{fsd50k}
Eduardo Fonseca, Xavier Favory, Jordi Pons, Frederic Font, and Xavier Serra.
\newblock {FSD50K}: An open dataset of human-labeled sound events.
\newblock \emph{IEEE/ACM Transactions on Audio, Speech, and Language Processing}, 30:\penalty0 829--852, 2022.
\newblock \doi{10.1109/TASLP.2021.3133208}.

\bibitem[Gayer(2026)]{shroom}
Yhonatan Gayer.
\newblock {SHroom}: A python framework for ambisonics room acoustics simulation and binaural rendering, 2026.
\newblock URL \url{https://arxiv.org/abs/2603.27342}.

\bibitem[Gayer et~al.(2025)Gayer, Tourbabin, Ben-Hur, Alon, and Rafaely]{gayer2025ambisonicsencoderwearablearray}
Yhonatan Gayer, Vladimir Tourbabin, Zamir Ben-Hur, David Alon, and Boaz Rafaely.
\newblock Ambisonics encoder for wearable array with improved binaural reproduction, 2025.
\newblock URL \url{https://arxiv.org/abs/2507.04108}.

\bibitem[Gemmeke et~al.(2017)Gemmeke, Ellis, Freedman, Jansen, Lawrence, Moore, Plakal, and Ritter]{audioset}
Jort~F. Gemmeke, Daniel P.~W. Ellis, Dylan Freedman, Aren Jansen, Wade Lawrence, R.~Channing Moore, Manoj Plakal, and Marvin Ritter.
\newblock {Audio Set}: An ontology and human-labeled dataset for audio events.
\newblock In \emph{2017 IEEE International Conference on Acoustics, Speech and Signal Processing (ICASSP)}, pages 776--780, 2017.
\newblock \doi{10.1109/ICASSP.2017.7952261}.

\bibitem[Golub et~al.(1999)Golub, Hansen, and O'Leary]{tik}
Gene~H. Golub, Per~Christian Hansen, and Dianne~P. O'Leary.
\newblock Tikhonov regularization and total least squares.
\newblock \emph{SIAM Journal on Matrix Analysis and Applications}, 21\penalty0 (1):\penalty0 185--194, 1999.
\newblock \doi{10.1137/S0895479897326432}.
\newblock URL \url{https://doi.org/10.1137/S0895479897326432}.

\bibitem[Gong et~al.(2022)Gong, Lai, Chung, and Glass]{gong2022ssast}
Yuan Gong, Cheng-I Lai, Yu-An Chung, and James Glass.
\newblock {SSAST}: Self-supervised audio spectrogram transformer.
\newblock In \emph{Proceedings of the AAAI Conference on Artificial Intelligence}, volume~36, pages 10699--10709, 2022.

\bibitem[Gu and Dao(2023)]{mamba}
Albert Gu and Tri Dao.
\newblock Mamba: Linear-time sequence modeling with selective state spaces.
\newblock \emph{arXiv preprint arXiv:2312.00752}, 2023.

\bibitem[Heikkinen et~al.(2026)Heikkinen, Politis, Drossos, and Virtanen]{heikkinen2026omnidirectionalneuralambisonicsencoding}
Mikko Heikkinen, Archontis Politis, Konstantinos Drossos, and Tuomas Virtanen.
\newblock Beyond omnidirectional: Neural ambisonics encoding for arbitrary microphone directivity patterns using cross-attention, 2026.
\newblock URL \url{https://arxiv.org/abs/2601.23196}.

\bibitem[Hsu et~al.(2021)Hsu, Bolte, Tsai, Lakhotia, Salakhutdinov, and Mohamed]{hubert}
Wei-Ning Hsu, Benjamin Bolte, Yao-Hung~Hubert Tsai, Kushal Lakhotia, Ruslan Salakhutdinov, and Abdelrahman Mohamed.
\newblock {HuBERT}: Self-supervised speech representation learning by masked prediction of hidden units.
\newblock \emph{IEEE/ACM Trans. Audio, Speech and Lang. Proc.}, 29:\penalty0 3451–3460, October 2021.
\newblock ISSN 2329-9290.
\newblock \doi{10.1109/TASLP.2021.3122291}.
\newblock URL \url{https://doi.org/10.1109/TASLP.2021.3122291}.

\bibitem[Huang et~al.(2022)Huang, Xu, Li, Baevski, Auli, Galuba, Metze, and Feichtenhofer]{audiomae}
Po-Yao Huang, Hu~Xu, Juncheng Li, Alexei Baevski, Michael Auli, Wojciech Galuba, Florian Metze, and Christoph Feichtenhofer.
\newblock Masked autoencoders that listen.
\newblock In \emph{NeurIPS}, 2022.

\bibitem[Jiang et~al.(2025)Jiang, Gamper, and Braun]{sciphi}
Xilin Jiang, Hannes Gamper, and Sebastian Braun.
\newblock Sci-phi: A large language model spatial audio descriptor, 2025.
\newblock URL \url{https://arxiv.org/abs/2510.05542}.

\bibitem[Kim et~al.(2025)Kim, Yun, and Kim]{kim2025visage}
Jaeyeon Kim, Heeseung Yun, and Gunhee Kim.
\newblock {ViSAGe}: Video-to-spatial audio generation.
\newblock In \emph{ICLR}, 2025.

\bibitem[Li et~al.(2024)Li, Shao, and Li]{atst}
Xian Li, Nian Shao, and Xiaofei Li.
\newblock Self-supervised audio teacher-student transformer for both clip-level and frame-level tasks.
\newblock \emph{IEEE/ACM Transactions on Audio, Speech, and Language Processing}, 32:\penalty0 1336--1351, 2024.
\newblock \doi{10.1109/TASLP.2024.3352248}.

\bibitem[Loshchilov and Hutter(2017)]{loshchilov2017decoupled}
Ilya Loshchilov and Frank Hutter.
\newblock Decoupled weight decay regularization.
\newblock \emph{arXiv preprint arXiv:1711.05101}, 2017.

\bibitem[Maciejewski et~al.(2020)Maciejewski, Wichern, and Le~Roux]{Maciejewski2020WHAMR}
Matthew Maciejewski, Gordon Wichern, and Jonathan Le~Roux.
\newblock {WHAMR}!: Noisy and reverberant single-channel speech separation.
\newblock In \emph{Proc. IEEE International Conference on Acoustics, Speech and Signal Processing (ICASSP)}, May 2020.

\bibitem[McCormack et~al.(2022)McCormack, Politis, Gonzalez, Lokki, and Pulkki]{acm_ref}
Leo McCormack, Archontis Politis, Raimundo Gonzalez, Tapio Lokki, and Ville Pulkki.
\newblock Parametric ambisonic encoding of arbitrary microphone arrays.
\newblock \emph{IEEE/ACM Transactions on Audio, Speech, and Language Processing}, 30:\penalty0 2062--2075, 2022.
\newblock \doi{10.1109/TASLP.2022.3182857}.

\bibitem[Mesaros et~al.(2018)Mesaros, Heittola, Benetos, Foster, Lagrange, Virtanen, and Plumbley]{Mesaros2018_TASLP}
A.~Mesaros, T.~Heittola, E.~Benetos, P.~Foster, M.~Lagrange, T.~Virtanen, and M.~D. Plumbley.
\newblock Detection and classification of acoustic scenes and events: Outcome of the {DCASE} 2016 challenge.
\newblock \emph{IEEE/ACM Transactions on Audio, Speech, and Language Processing}, 26\penalty0 (2):\penalty0 379--393, Feb 2018.
\newblock ISSN 2329-9290.
\newblock \doi{10.1109/TASLP.2017.2778423}.

\bibitem[Morgado et~al.(2020)Morgado, Li, and Nvasconcelos]{yt360}
Pedro Morgado, Yi~Li, and Nuno Nvasconcelos.
\newblock Learning representations from audio-visual spatial alignment.
\newblock \emph{Advances in Neural Information Processing Systems}, 33, 2020.

\bibitem[Niizumi et~al.(2022)Niizumi, Takeuchi, Ohishi, Harada, and Kashino]{masked-modelling}
Daisuke Niizumi, Daiki Takeuchi, Yasunori Ohishi, Noboru Harada, and Kunio Kashino.
\newblock Masked spectrogram modeling using masked autoencoders for learning general-purpose audio representation.
\newblock In Joseph Turian, Björn~W. Schuller, Dorien Herremans, Katrin Kirchoff, Paola~Garcia Perera, and Philippe Esling, editors, \emph{HEAR: Holistic Evaluation of Audio Representations (NeurIPS 2021 Competition)}, volume 166 of \emph{Proceedings of Machine Learning Research}, pages 1--24. PMLR, 13--14 Dec 2022.
\newblock URL \url{https://proceedings.mlr.press/v166/niizumi22a.html}.

\bibitem[Panayotov et~al.(2015)Panayotov, Chen, Povey, and Khudanpur]{librispeech}
Vassil Panayotov, Guoguo Chen, Daniel Povey, and Sanjeev Khudanpur.
\newblock Librispeech: An asr corpus based on public domain audio books.
\newblock In \emph{2015 IEEE International Conference on Acoustics, Speech and Signal Processing (ICASSP)}, pages 5206--5210, 2015.
\newblock \doi{10.1109/ICASSP.2015.7178964}.

\bibitem[Piczak()]{esc50}
Karol~J. Piczak.
\newblock {ESC}: {Dataset} for {Environmental Sound Classification}.
\newblock In \emph{Proceedings of the 23rd {Annual ACM Conference} on {Multimedia}}, pages 1015--1018. {ACM Press}.
\newblock ISBN 978-1-4503-3459-4.
\newblock \doi{10.1145/2733373.2806390}.
\newblock URL \url{http://dl.acm.org/citation.cfm?doid=2733373.2806390}.

\bibitem[Politis(2022)]{politis_2022_6406873}
Archontis Politis.
\newblock [dcase2022 task 3] synthetic seld mixtures for baseline training, April 2022.
\newblock URL \url{https://doi.org/10.5281/zenodo.6406873}.

\bibitem[Politis et~al.(2020{\natexlab{a}})Politis, Adavanne, and Virtanen]{tau2020}
Archontis Politis, Sharath Adavanne, and Tuomas Virtanen.
\newblock A dataset of reverberant spatial sound scenes with moving sources for sound event localization and detection.
\newblock In \emph{Proceedings of the Detection and Classification of Acoustic Scenes and Events 2020 Workshop ({DCASE2020})}, pages 165--169, Tokyo, Japan, November 2020{\natexlab{a}}.
\newblock URL \url{https://dcase.community/workshop2020/proceedings}.

\bibitem[Politis et~al.(2020{\natexlab{b}})Politis, Mesaros, Adavanne, Heittola, and Virtanen]{eval_overview}
Archontis Politis, Annamaria Mesaros, Sharath Adavanne, Toni Heittola, and Tuomas Virtanen.
\newblock Overview and evaluation of sound event localization and detection in {DCASE 2019}.
\newblock \emph{IEEE/ACM Transactions on Audio, Speech, and Language Processing}, 29:\penalty0 684--698, 2020{\natexlab{b}}.
\newblock URL \url{https://ieeexplore.ieee.org/abstract/document/9306885}.

\bibitem[Politis et~al.(2021)Politis, Adavanne, Krause, Deleforge, Srivastava, and Virtanen]{tau2021}
Archontis Politis, Sharath Adavanne, Daniel Krause, Antoine Deleforge, Prerak Srivastava, and Tuomas Virtanen.
\newblock A dataset of dynamic reverberant sound scenes with directional interferers for sound event localization and detection.
\newblock In \emph{Proceedings of the 6th Detection and Classification of Acoustic Scenes and Events 2021 Workshop ({DCASE2021})}, pages 125--129, Barcelona, Spain, November 2021.
\newblock ISBN 978-84-09-36072-7.
\newblock URL \url{https://dcase.community/workshop2021/proceedings}.

\bibitem[Politis et~al.(2022)Politis, Shimada, Sudarsanam, Adavanne, Krause, Koyama, Takahashi, Takahashi, Mitsufuji, and Virtanen]{STARSS22}
Archontis Politis, Kazuki Shimada, Parthasaarathy Sudarsanam, Sharath Adavanne, Daniel Krause, Yuichiro Koyama, Naoya Takahashi, Shusuke Takahashi, Yuki Mitsufuji, and Tuomas Virtanen.
\newblock {STARSS22}: A dataset of spatial recordings of real scenes with spatiotemporal annotations of sound events, 2022.
\newblock URL \url{https://arxiv.org/abs/2206.01948}.

\bibitem[Saeed et~al.(2021)Saeed, Grangier, and Zeghidour]{COLA}
Aaqib Saeed, David Grangier, and Neil Zeghidour.
\newblock Contrastive learning of general-purpose audio representations.
\newblock In \emph{ICASSP 2021 - 2021 IEEE International Conference on Acoustics, Speech and Signal Processing (ICASSP)}, pages 3875--3879, 2021.
\newblock \doi{10.1109/ICASSP39728.2021.9413528}.

\bibitem[Sarabia et~al.(2023)Sarabia, Menyaylenko, Toso, Seto, Aldeneh, Pirhosseinloo, Zappella, Theobald, Apostoloff, and Sheaffer]{spatial-libri}
Miguel Sarabia, Elena Menyaylenko, Alessandro Toso, Skyler Seto, Zakaria Aldeneh, Shadi Pirhosseinloo, Luca Zappella, Barry-John Theobald, Nicholas Apostoloff, and Jonathan Sheaffer.
\newblock {Spatial LibriSpeech: An Augmented Dataset for Spatial Audio Learning}.
\newblock In \emph{Proc. Interspeech}, pages 3724--3728, 2023.
\newblock \doi{10.21437/Interspeech.2023-2117}.

\bibitem[Schmidt(1986)]{MUSIC}
R.~Schmidt.
\newblock Multiple emitter location and signal parameter estimation.
\newblock \emph{IEEE Transactions on Antennas and Propagation}, 34\penalty0 (3):\penalty0 276--280, 1986.
\newblock \doi{10.1109/TAP.1986.1143830}.

\bibitem[Sheelvant et~al.(2019)Sheelvant, Sharma, Madhavi, Das, Prasanna, and Li]{rsl2019}
Rohan Sheelvant, Bidisha Sharma, Maulik Madhavi, Rohan~Kumar Das, S.~R.~M. Prasanna, and Haizhou Li.
\newblock {RSL2019}: A realistic speech localization corpus.
\newblock In \emph{2019 22nd Conference of the Oriental COCOSDA International Committee for the Co-ordination and Standardisation of Speech Databases and Assessment Techniques (O-COCOSDA)}, pages 1--6, 2019.
\newblock \doi{10.1109/O-COCOSDA46868.2019.9060842}.

\bibitem[Shimada et~al.(2021)Shimada, Koyama, Takahashi, Takahashi, and Mitsufuji]{shimada2021accdoa}
Kazuki Shimada, Yuichiro Koyama, Naoya Takahashi, Shusuke Takahashi, and Yuki Mitsufuji.
\newblock {ACCDOA}: Activity-coupled cartesian direction of arrival representation for sound event localization and detection, 2021.
\newblock URL \url{https://arxiv.org/abs/2010.15306}.

\bibitem[Shimada et~al.(2023)Shimada, Politis, Sudarsanam, Krause, Uchida, Adavanne, Hakala, Koyama, Takahashi, Takahashi, Virtanen, and Mitsufuji]{STARSS23}
Kazuki Shimada, Archontis Politis, Parthasaarathy Sudarsanam, Daniel Krause, Kengo Uchida, Sharath Adavanne, Aapo Hakala, Yuichiro Koyama, Naoya Takahashi, Shusuke Takahashi, Tuomas Virtanen, and Yuki Mitsufuji.
\newblock {STARSS23}: an audio-visual dataset of spatial recordings of real scenes with spatiotemporal annotations of sound events.
\newblock In \emph{Proceedings of the 37th International Conference on Neural Information Processing Systems}, NIPS '23, Red Hook, NY, USA, 2023. Curran Associates Inc.

\bibitem[Stöter et~al.(2018)Stöter, Chakrabarty, Habets, and Edler]{libricount}
Fabian-Robert Stöter, Soumitro Chakrabarty, Emanuël Habets, and Bernd Edler.
\newblock Libricount, a dataset for speaker count estimation, April 2018.
\newblock URL \url{https://doi.org/10.5281/zenodo.1216072}.

\bibitem[Tang et~al.(2024)Tang, Yu, Sun, Chen, Tan, Li, Zhang, Lu, Ma, Wang, and Zhang]{tang2024largelanguagemodelsunderstand}
Changli Tang, Wenyi Yu, Guangzhi Sun, Xianzhao Chen, Tian Tan, Wei Li, Jun Zhang, Lu~Lu, Zejun Ma, Yuxuan Wang, and Chao Zhang.
\newblock Can large language models understand spatial audio?, 2024.
\newblock URL \url{https://arxiv.org/abs/2406.07914}.

\bibitem[Tian et~al.(2014)Tian, Srinivasamurthy, Sandler, and Serra]{Beijingop}
Mi~Tian, Ajay Srinivasamurthy, Mark Sandler, and Xavier Serra.
\newblock A study of instrument-wise onset detection in {Beijing Opera} percussion ensembles.
\newblock In \emph{2014 IEEE International Conference on Acoustics, Speech and Signal Processing (ICASSP)}, pages 2159--2163, 2014.
\newblock \doi{10.1109/ICASSP.2014.6853981}.

\bibitem[Turian et~al.(2022)Turian, Shier, Khan, Raj, Schuller, Steinmetz, Malloy, Tzanetakis, Velarde, McNally, Henry, Pinto, Noufi, Clough, Herremans, Fonseca, Engel, Salamon, Esling, Manocha, Watanabe, Jin, and Bisk]{hear}
Joseph Turian, Jordie Shier, Humair~Raj Khan, Bhiksha Raj, Bj\"{o}rn~W. Schuller, Christian~J. Steinmetz, Colin Malloy, George Tzanetakis, Gissel Velarde, Kirk McNally, Max Henry, Nicolas Pinto, Camille Noufi, Christian Clough, Dorien Herremans, Eduardo Fonseca, Jesse Engel, Justin Salamon, Philippe Esling, Pranay Manocha, Shinji Watanabe, Zeyu Jin, and Yonatan Bisk.
\newblock {HEAR: Holistic Evaluation of Audio Representations}.
\newblock In Douwe Kiela, Marco Ciccone, and Barbara Caputo, editors, \emph{Proceedings of the NeurIPS 2021 Competitions and Demonstrations Track}, volume 176 of \emph{Proceedings of Machine Learning Research}, pages 125--145. PMLR, 06--14 Dec 2022.
\newblock URL \url{https://proceedings.mlr.press/v176/turian22a.html}.

\bibitem[Valk and Alumäe(2021)]{voxling}
Jörgen Valk and Tanel Alumäe.
\newblock Voxlingua107: A dataset for spoken language recognition.
\newblock In \emph{2021 IEEE Spoken Language Technology Workshop (SLT)}, pages 652--658, 2021.
\newblock \doi{10.1109/SLT48900.2021.9383459}.

\bibitem[van~der Heijden et~al.(2019)van~der Heijden, Rauschecker, de~Gelder, and Formisano]{heijden2019cortical}
Kiki van~der Heijden, Josef~P Rauschecker, Beatrice de~Gelder, and Elia Formisano.
\newblock Cortical mechanisms of spatial hearing.
\newblock \emph{Nature Reviews Neuroscience}, 20\penalty0 (10):\penalty0 609--623, 2019.

\bibitem[Wang et~al.(2022)Wang, Politis, Mesaros, and Virtanen]{ambisonic_IVs}
Shanshan Wang, Archontis Politis, Annamaria Mesaros, and Tuomas Virtanen.
\newblock Self-supervised learning of audio representations from audio-visual data using spatial alignment.
\newblock \emph{IEEE Journal of Selected Topics in Signal Processing}, 16\penalty0 (6):\penalty0 1467--1479, 2022.
\newblock \doi{10.1109/JSTSP.2022.3180592}.

\bibitem[Warden(2018)]{warden2018speechcommandsdatasetlimitedvocabulary}
Pete Warden.
\newblock Speech commands: A dataset for limited-vocabulary speech recognition, 2018.
\newblock URL \url{https://arxiv.org/abs/1804.03209}.

\bibitem[Yadav and Tan(2024)]{audiomamba}
Sarthak Yadav and Zheng-Hua Tan.
\newblock {Audio Mamba}: Selective state spaces for self-supervised audio representations.
\newblock In \emph{Interspeech 2024}, pages 552--556, 2024.
\newblock \doi{10.21437/Interspeech.2024-1274}.

\bibitem[Yadav et~al.(2024)Yadav, Theodoridis, Hansen, and Tan]{mwmae}
Sarthak Yadav, Sergios Theodoridis, Lars~Kai Hansen, and Zheng-Hua Tan.
\newblock Masked autoencoders with multi-window local-global attention are better audio learners.
\newblock In \emph{The Twelfth International Conference on Learning Representations}, 2024.
\newblock URL \url{https://openreview.net/forum?id=Q53QLftNkA}.

\bibitem[Yang et~al.(2021)Yang, Chi, Chuang, Lai, Lakhotia, Lin, Liu, Shi, Chang, Lin, Huang, Tseng, tik Lee, Liu, Huang, Dong, Li, Watanabe, Mohamed, and yi~Lee]{superb}
Shuwen Yang, Po-Han Chi, Yung-Sung Chuang, Cheng-I~Jeff Lai, Kushal Lakhotia, Yist~Y. Lin, Andy~T. Liu, Jiatong Shi, Xuankai Chang, Guan-Ting Lin, Tzu-Hsien Huang, Wei-Cheng Tseng, Ko~tik Lee, Da-Rong Liu, Zili Huang, Shuyan Dong, Shang-Wen Li, Shinji Watanabe, Abdelrahman Mohamed, and Hung yi~Lee.
\newblock Superb: Speech processing universal performance benchmark.
\newblock In \emph{Interspeech 2021}, pages 1194--1198, 2021.
\newblock \doi{10.21437/Interspeech.2021-1775}.

\bibitem[Zheng et~al.(2024)Zheng, Peng, Ma, Chen, Choi, and Harwath]{spatialast}
Zhisheng Zheng, Puyuan Peng, Ziyang Ma, Xie Chen, Eunsol Choi, and David Harwath.
\newblock Bat: learning to reason about spatial sounds with large language models.
\newblock In \emph{Proceedings of the 41st International Conference on Machine Learning}, ICML'24. JMLR.org, 2024.

\bibitem[Zotter and Frank(2019)]{ambisonics}
Franz Zotter and Matthias Frank.
\newblock \emph{XY, MS, and First-Order Ambisonics}, pages 1--22.
\newblock Springer International Publishing, Cham, 2019.
\newblock ISBN 978-3-030-17207-7.
\newblock \doi{10.1007/978-3-030-17207-7_1}.
\newblock URL \url{https://doi.org/10.1007/978-3-030-17207-7_1}.

\end{thebibliography}
